\documentclass[twocolumn,trackchanges]{aastex701}

\usepackage{comment}
\usepackage{rotating}
\usepackage{ulem}
\usepackage{soul}
\usepackage{multirow}
\usepackage{longtable}
\usepackage{indentfirst}  
\usepackage{longtable}  
\usepackage{tablefootnote}
\usepackage{amsmath}
\shorttitle{Pulsars in M92}
\shortauthors{Yin et al.}

\received{?}
\revised{?}
\accepted{?}

\submitjournal{ApJL}

\graphicspath{{./}{figures/}}

\begin{document}

\title{From One to Two: A Second Binary Millisecond Pulsar in the Globular Cluster M92 (NGC~6341)}
% Discovery of a Binary Millisecond Pulsar in the Globular Cluster M92 (NGC 6341)
% Discovery of a Second Millisecond Pulsar Associated with the Globular Cluster M92 (NGC 6341)
% Doubling the Pulsars in M92: Discovery of a Binary Millisecond Pulsar
% From One to Two: A Second Binary Millisecond Pulsar in the Globular Cluster M92 (NGC 6341)
\author[0000-0001-6051-3420]{Dejiang Yin}
\affiliation{College of Physics, Guizhou University, Guiyang 550025, People's Republic of China}
%\email[show]{dj.yin@foxmail.com}
\email{dj.yin@foxmail.com}

\author[0000-0002-2394-9521]{Li-yun Zhang}
\affiliation{College of Physics, Guizhou University, Guiyang 550025, People's Republic of China}
\affiliation{Guizhou Radio Astronomical Observatory, Guizhou University, Guiyang 550025, People's Republic of China}
\affiliation{International Centre of Supernovae, Yunnan Key Laboratory, Kunming 650216, People's Republic of China}
\email[show]{liy\_zhang@hotmail.com}

\author[0009-0008-4109-744X]{Baoda Li}
\affiliation{College of Physics, Guizhou University, Guiyang 550025, People's Republic of China}
\email{baodali@foxmail.com}

\author[0009-0007-6396-7891]{Yinfeng Dai}
\affiliation{Institute for Frontier in Astronomy and Astrophysics \& Faculty of Arts and Sciences, Beijing Normal University, Zhuhai 519087, People's Republic of China}
\affiliation{School of Physics and Astronomy, Beijing Normal University, Beijing 100875, People's Republic of China}
\email{yfeng.dai@foxmail.com}

\author[0000-0003-0757-3584]{Lin Wang}
\affiliation{State Key Laboratory of Radio Astronomy and Technology, Shanghai Astronomical Observatory, Chinese Academy of Sciences, 80 Nandan Road, Shanghai 200030, People's Republic of China}
% \affiliation{Kavli Institute for Astronomy and Astrophysics, Peking University, Beijing 100871, People's Republic of China}
\email{wanglin@shao.ac.cn}

\author[0009-0002-2059-6350]{Qiuyu Yu}
\affiliation{College of Physics, Guizhou University, Guiyang 550025, People's Republic of China}
\affiliation{Guizhou Vocational College of Foodstuff Engineering, Guiyang 550025, People's Republic of China}
\email{gs.qyyu23@gzu.edu.cn}

\author[0009-0001-6693-7555]{Yujie Lian} 
\affiliation{School of Physics and Astronomy, Beijing Normal University, Beijing 100875, People's Republic of China}
\email{yujielian@mail.bnu.edu.cn}

\begin{abstract}
We report the discovery and phase connected-timing solution of a second millisecond binary pulsar, PSR~J1717+4308B (M92B), in the globular cluster M92 (NGC~6341) using the Five-hundred-meter Aperture Spherical radio Telescope.
This new pulsar, with a spin period of 3.51\,ms and a dispersion measure (DM) of 35.29\,pc\,cm$^{-3}$, was discovered through frequency-domain acceleration searches.
The timing solution shows that M92B is in a binary system with an orbital period of 2.3\,days, an eccentricity of $\simeq 4.8 \times 10^{-4}$, and a minimum companion mass of 0.2~$\,M_\odot$.
M92B lies within the cluster core radius in projection, and its negative spin period derivative ($\dot{P}$) is consistent with acceleration in the cluster potential.
The measured negative $\dot{P}$ of M92B, together with a DM consistent with that of M92A ($< 0.2\, \rm pc\,cm^{-3}$), confirms that both pulsars are members of the cluster.
A Bayesian Markov Chain Monte Carlo analysis based on these two pulsars yields broad constraints on the core structural parameters of M92 that are consistent with \(N\)-body dynamical modeling. This demonstrates that pulsar timing can provide useful dynamical information in sparse pulsar samples.

\end{abstract}
\keywords{Globular star clusters (656); Millisecond pulsars (1062); Radio telescopes (1360)}

\section{Introduction}
\label{Intro}

Globular clusters (GCs) are favorable environments for producing millisecond pulsars (MSPs) owing to their extremely high stellar densities ($\gtrsim \, 10^3~ M_{\odot}\rm \, pc^{-3}$; \citealt{2010arXiv1012.3224H}) compared to the Galactic field \citep{1975ApJ...199L.143C, 1982Natur.300..728A}.
Pulsars in GCs\footnote{See the continuously updated catalog of GC pulsars maintained by P. C. C. Freire: \url{https://www3.mpifr-bonn.mpg.de/staff/pfreire/GCpsr.html}.} include extreme systems such as the fastest-spinning pulsar \citep{2006Sci...311.1901H} and candidate pulsar-black hole binaries \citep{2024Sci...383..275B}, which are exclusive products of cluster dynamics and are rarely found in the Galactic disk. 
Beyond the study of individual exotic systems, a population of pulsars within a cluster can probe the properties of its host cluster, e.g., the gravitational potential and possible central intermediate-mass black holes (e.g., \citealt{2017MNRAS.471..857F, 2017MNRAS.468.2114P, 2024ApJ...972..198C}), and intracluster magnetic field and gas content \citep{2001ApJ...557L.105F, 2018MNRAS.481..627A}.
Accordingly, given their rich scientific potential, GCs have long been prime targets for pulsar searches and studies
(e.g., \citealt{2000ApJ...535..975C, 2005Sci...307..892R, 2007ApJ...670..363H,  2021MNRAS.504.1407R, 2021ApJ...915L..28P}).

GC pulsars are typically faint radio sources located at significant distance (e.g., $>$2\,kpc, \citealt{2010arXiv1012.3224H}), resulting in low observed flux densities (e.g., \citealt{2007ApJ...670..363H, 2021MNRAS.504.1407R, 2021ApJ...915L..28P, 2025ApJS..279...51L}).
Most GC pulsars ($\sim 93\%$ of the total) are MSPs, and about 56\% of the population reside in binary systems \citep{2008IAUS..246..291R, 2023RAA....23e5012Y}. 
Orbital motion in binary systems 
modulates the pulsar signal, spreading the signal power over a range of frequencies and reducing the detectability of faint signals \citep{2004hpa..book.....L}.
However, interstellar scintillation can occasionally enhance the apparent flux density of pulsars above the detection threshold, enabling the detection of previously missed pulsars that were below survey sensitivity limits (e.g., M13G--I; \citealt{2025ApJ...991...38L, 2025ApJ...991..177Y}).

Empirical estimates of GC pulsar populations suggest that M92 is among the most promising targets for deep searches (see \citealt{2016RAA....16..151Z, 2020ApJ...892L...6P}) with the Five-hundred-meter Aperture Spherical radio Telescope (FAST; \citealt{2011IJMPD..20..989N, 2020RAA....20...64J}). 
The first pulsar in M92, PSR~J1717+4308A (M92A), was discovered during an early observation of GC FANS\footnote{Globular Cluster with FAST: A Neutron-star Survey; \citealt{2021ApJ...915L..28P,2025ApJS..279...51L}} on 2017 October 9 \citep{2020ApJ...892L...6P}.
M92A is a 3.15\,ms MSP in an eclipsing ``redback'' binary with an orbital period of 4.8\,hr. 
Its $X$-ray counterpart was identified and exhibits a luminosity of $8.3\times10^{31}$\,erg\,s$^{-1}$ in the 0.3--8 keV band from $Chandra$ observation \citep{2022MNRAS.511.5964Z}.
The first phase-connected timing solution for M92A was obtained from timing observations in 2021 \citep{2021ApJ...915L..28P} and recently updated in 2025 \citep{2025ApJS..279...51L}.
These data also provide an opportunity to search for additional pulsars that may reside within the cluster (e.g., \citealt{2013MNRAS.436.3720T, 2024ApJ...969L...7Y}), whose discovery would provide further evidence for the association of M92A with M92 \citep{2020ApJ...892L...6P}.

In this Letter, we searched for additional pulsars in M92 using 6\,yr of FAST timing observations for M92A. This extensive baseline significantly increased the chances of detecting faint pulsars that may occasionally become detectable due to interstellar scintillation. 
The structure of this Letter is as follows. Section~\ref{sec:Obs-Data} describes the FAST observations and data reduction. Section~\ref{sec:results} presents the discovery and timing results of the new pulsar. Section~\ref{discussion} and Section~\ref{Conclusions} provide a brief discussion and summary.

\section{Observation and data reduction} \label{sec:Obs-Data}
\subsection{FAST Observation} \label{sec:floats}

The early FAST observations of M92, together with the observing setup and initial data reduction, were described by \citet{2020ApJ...892L...6P, 2021ApJ...915L..28P}. Here we analyze 25 observations obtained between 2019 June 25 and 2025 August 26.
The integration times range from 1800 to 18,000\,s.
M92 has a core radius of $r_{\rm c}=0.26'$ and a half-light radius of $r_{\rm h}=1.02'$ \citep[]{2010arXiv1012.3224H}. Both are well within the $2.9'$ half-power beamwidth of the FAST 19-beam $L$-band receiver.
The central beam was pointed at the cluster center (R.A. = 17:17:07.39, decl. = +43:08:09.4; \citealt{2010arXiv1012.3224H}) for all observations, and only the central-beam data were analyzed in this work.
All data were taken
over 1.0--1.5\,GHz, channelized into 4096 channels of width 0.122\,MHz \citep{2019SCPMA..6259502J}. 
These observations were carried out in the \texttt{Tracking} or \texttt{SwiftCalibration} mode with a sampling time of $49.152\,\mu{\rm s}$ and recorded in the standard search-mode PSRFITS format \citep{2004PASA...21..302H}.

\subsection{Data Reduction} \label{sec:floats}
 
Pulsar searches were performed using the \textsc{PulsaR Exploration and Search TOolkit} (\textsc{PRESTO}\footnote{\url{https://github.com/scottransom/presto}}; \citealt{2001PhDT.......123R,2002AJ....124.1788R}) package with the \textsc{MOSS} scripts\footnote{Multiple Observation Segment Search (MOSS) for Pulsars: \url{https://github.com/ydejiang/MOSS}; see details in  \citealt{2025ApJ...991..177Y}} \citep{dejiang_yin_2025_20178378}. 
Radio-frequency interference was mitigated in both the time and frequency domains using the \texttt{rfifind} routine. Dedispersed time series were then generated from the PSRFITS data using \texttt{prepsubband} routine over a range of trial dispersion measures (DMs).
The previously known pulsar in this cluster, M92A, has a DM of $35.45~{\rm pc~cm^{-3}}$ \citep{2020ApJ...892L...6P}. 
Following the empirical relation between the mean DM and DM spread of pulsars in GCs \citep{2023RAA....23e5012Y}, we searched a DM range broader than the expected spread around the known pulsar M92A ($35.45~{\rm pc~cm^{-3}}$), covering $32$–$38~{\rm pc~cm^{-3}}$ with a step size of $0.05~{\rm pc~cm^{-3}}$.
This DM step corresponds to a pulse broadening of only $\sim2\%$ for a minimum spin period of $0.5$~ms considered in our search \citep{2007ApJ...670..363H}.

The dedispersed time series were transformed to the fluctuation frequency domain using the \texttt{realfft} routine, and low-frequency red noise was mitigated with the \texttt{rednoise} routine. 
Fourier-domain acceleration searches were then performed with the \texttt{accelsearch} routine \citep{2002AJ....124.1788R}.
To enhance sensitivity to short-orbit systems, observations longer than 3000\,s were divided into $\sim$1500--2500\,s segments for acceleration searches according to the observing duration at each epoch.
Acceleration searches were performed on both full-length observations and segmented data using harmonic summing up to 32. For computationally efficient searches of isolated pulsars and low-acceleration pulsars in long-orbit binaries, we used \texttt{$z_{\rm max}$} = 20. For segmented data targeting accelerated pulsars in short-orbit binaries, we initially used \texttt{$z_{\rm max}$} = 300, and later adopted \texttt{$z_{\rm max}$} = 520 after the new discovery to improve sensitivity to more accelerated signals.

We also performed ``jerk'' searches on the segmented data to account for rapidly varying accelerations.
Following the search setups explored by \citet{2018ApJ...863L..13A}, we used \texttt{$z_{\rm max}$} = 100 and \texttt{$w_{\rm max}$} = 500, as well as a wider search with \texttt{$z_{\rm max}$} = 300 and \texttt{$w_{\rm max}$} = 900, both with four-harmonic summing.

All candidates produced by the \texttt{accelsearch} routine were sifted using a customized version of the \texttt{ACCEL\_sift.py} script in \textsc{PRESTO}.
Each diagnostic plot used for initial inspection comprises two panels, showing the folded profile of the dedispersed time series (upper) and the trial DM-detected significance curve (lower).
Promising candidates were then refolded using the raw PSRFITS data to produce standard diagnostic plots for final verification.

Upon discovery of a binary pulsar,
initial orbital parameters were obtained using the \texttt{fitorb.py} script in \textsc{PRESTO}, based on the apparent barycentric spin periods and their derivatives measured at different orbital phases. 
Using initial parameters, the data were folded to construct a preliminary timing model.
Subsequent timing iterations refined the orbital solution, yielding more detections in observations where the pulsar signal is weak and affected by orbital modulation.
A standard template profile was constructed from high signal-to-noise detections using Gaussian-component fitting with \textsc{PRESTO}'s \texttt{pygaussfit.py} script. 
Pulse times of arrival (TOAs) were obtained with \texttt{get\_TOA.py} script using the Fourier-domain template matching technique \citep{1992RSPTA.341..117T}. 

\section{Results} \label{sec:results}
\subsection{Discovery and Timing of M92B}

\begin{figure}[h]
\begin{center}
\includegraphics[width=1\linewidth]{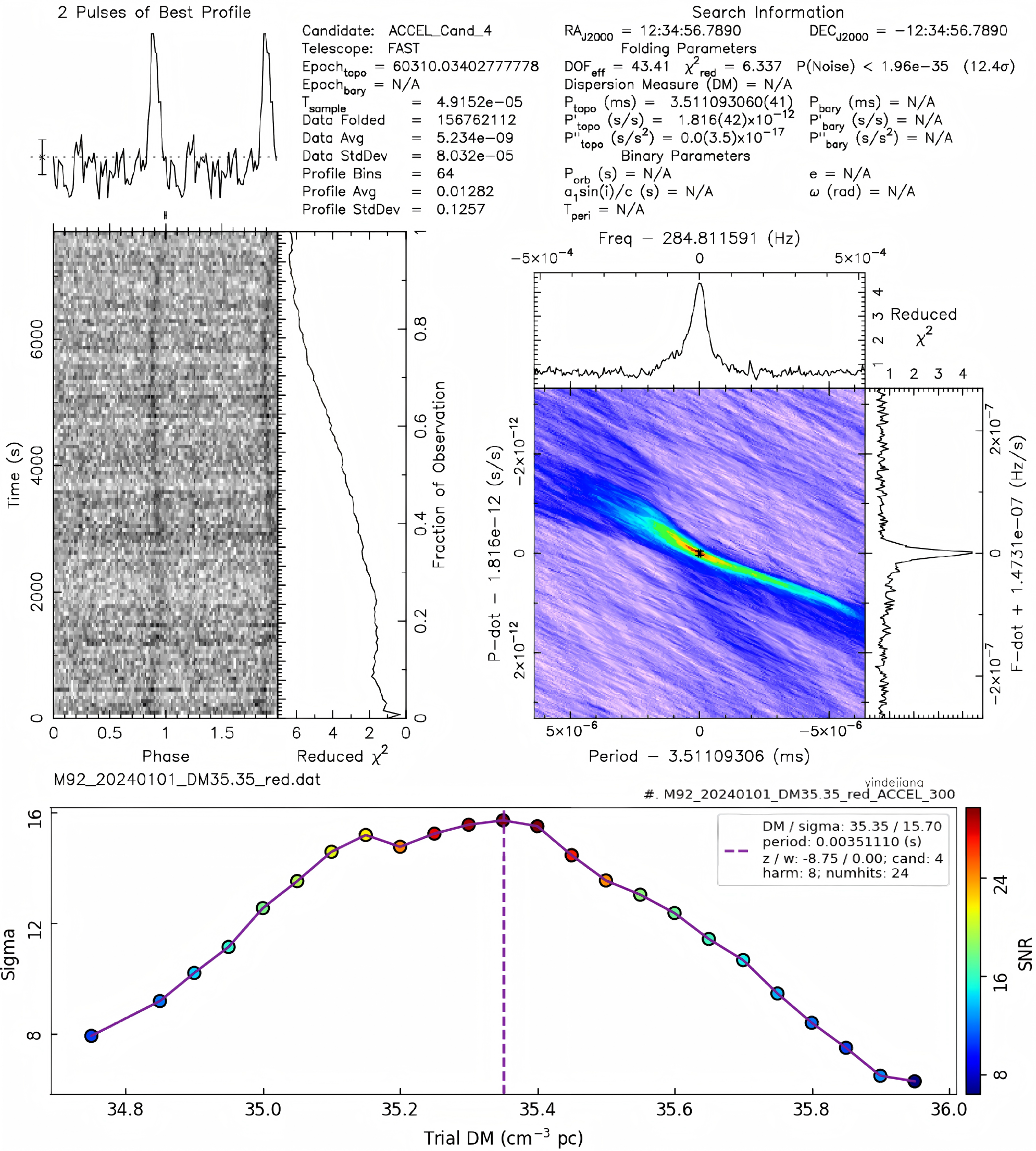}
\caption{
Discovery plot of M92B. Top: folded pulse profile from the dedispersed time series produced using \texttt{prepfold}. Bottom: detection significance as a function of trial DM, peaking near the optimal value and decreasing due to dispersive smearing.
}
\label{fig:M92B}
\end{center}
\end{figure}

\begin{figure*}[htp!]
\begin{center}
    \includegraphics[width=1.0\linewidth]{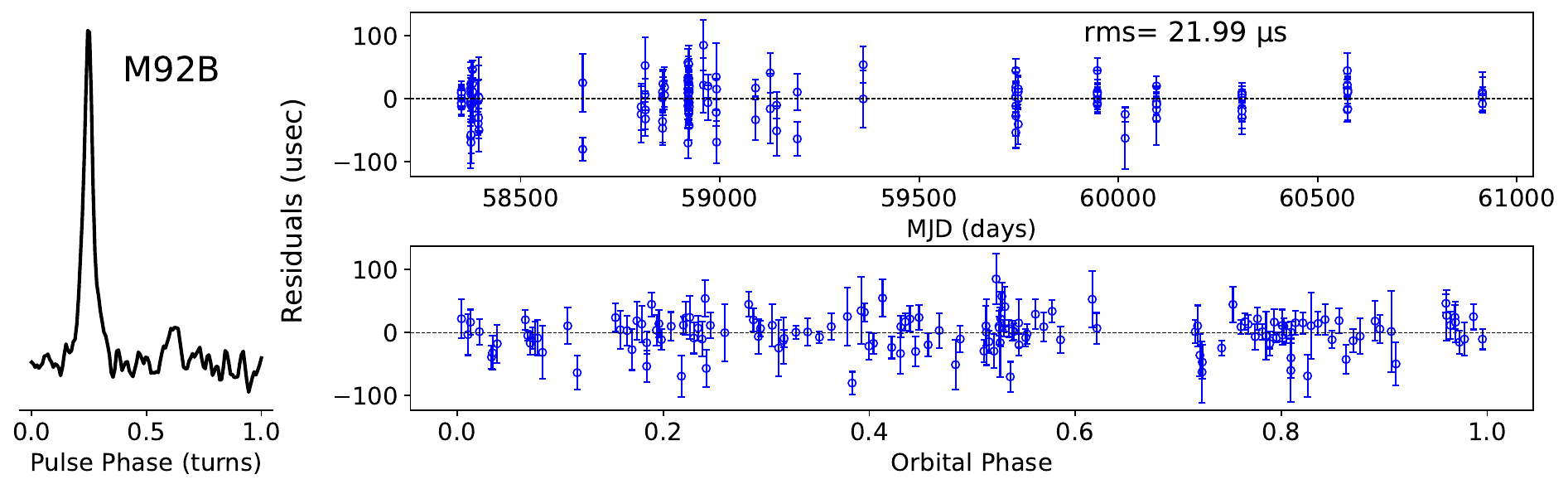}
\caption{
Average pulse profile and timing residuals of M92B. 
The left panel shows the integrated pulse profile obtained by summing all 36 detections over 128 phase bins. 
The right panel shows the timing residuals versus MJD (top) and orbital phase (bottom) from the best-fit timing model.
}\label{fig:timing}
\end{center}
\end{figure*}

Our acceleration searches (\texttt{$z_{\rm max}$} = 300 scheme) of the FAST monitoring observations revealed a new MSP in M92, PSR~J1717+4308B (hereafter M92B, see Figure~\ref{fig:M92B}). 
The pulsar was detected with a spin period of 3.51\,ms and a DM of 35.35\,pc\,cm$^{-3}$, consistent with the DM of the previously known pulsar M92A.
The discovery signal exhibited significant Doppler acceleration ($\sim 0.2\, \rm m\,s^{-2}$), suggesting that the pulsar is in a binary system.
Figure~\ref{fig:M92B} shows mild higher-order orbital effects in the acceleration-search discovery signal, corresponding to a small line-of-sight (LOS) jerk of $\sim 1.2 \times 10^{-5}\,\rm m\,s^{-3}$.
No eclipses were detected for M92B in our observations.

The timing analysis was performed using the \textsc{TEMPO}\footnote{\url{https://tempo.sourceforge.net/}} \citep{2015ascl.soft09002N} and \textsc{Dracula}\footnote{\url{https://github.com/pfreire163/Dracula}} \citep{2018MNRAS.476.4794F} packages, yielding a phase-connected timing solution with the ELL1 model \citep{2001MNRAS.326..274L}.
The timing results show that M92B is a binary MSP with an orbital period of 2.3\,days and a mildly eccentricity of \textbf{$e = 4.82(1)\times10^{-4}$. }
The derived timing parameters are listed in Table~\ref{table:timing}, and the corresponding timing residuals are shown in Figure~\ref{fig:timing}.

\begin{table}[htbp!]
\centering
\caption{The timing solution of M92B.}
\label{table:timing}

\footnotesize 
\setlength{\tabcolsep}{0pt} 

\begin{tabular*}{\columnwidth}{@{\extracolsep{\fill}}l r}
\hline
\hline
Pulsar & J1717+4308B \\
\hline\hline
Right Ascension, $\alpha$ (J2000) & 17:17:07.2234(1) \\
Declination, $\delta$ (J2000) & +43:08:07.890(1) \\
Spin Frequency, $f$ (s$^{-1}$) & 284.78215879637(4) \\
Spin Freq. deriv., $\dot{f}$ (s$^{-2}$) & 2.2093(5)$\times 10^{-15}$ \\
Reference Epoch (MJD) & 60310.032810 \\
Start of Timing Data (MJD) & 58351.448 \\
End of Timing Data (MJD) & 60913.479 \\
Dispersion Measure, DM (pc cm$^{-3}$) & 35.292(8) \\
Solar System Ephemeris & DE438 \\
Terrestrial Time Standard & UTC(NIST) \\
Time Units & TDB \\
Number of TOAs & 146 \\
Residuals RMS ($\mu$s) & 22.00 \\
\hline
\multicolumn{2}{c}{Binary Parameters} \\
\hline\hline
Binary Model & ELL1 \\
Proj. semi-major axis, $x_{\rm p}$ (lt-s) & 2.591407(3) \\
1st Laplace-Lagr. param., $\eta$ & 6.8(2)$\times 10^{-5}$ \\
2nd Laplace-Lagr. param., $\kappa$ & $-$4.78(2)$\times 10^{-4}$ \\
$T_\textrm{asc}$ (MJD) & 58920.9353449(4) \\
Orbital Period, $P_b$ (days) & 2.294099062(1) \\
\hline
\multicolumn{2}{c}{Derived Parameters} \\
\hline\hline
Spin Period, $P$ (s) & 3.5114559290740(4)$\times 10^{-3}$ \\
Spin Period deriv., $\dot{P}$ (s s$^{-1}$) & $-$2.7241(6)$\times 10^{-20}$ \\
Mass Function, $f(M_{\rm p})$ (${M}_\odot$) & 3.5502$\times 10^{-3}$ \\
Min. companion mass, $M_{\rm c, min}$ (${M}_\odot$) & 0.20 \\
Med. companion mass, $M_{\rm c, med}$ (${M}_\odot$) & 0.24 \\
Offset in $\alpha$, $\theta_\alpha$ (arcmin) & 0.0304 \\
Offset in $\delta$, $\theta_\delta$ (arcmin) & $-$0.0252 \\
Total offset, $\theta_\perp$ (arcmin) & 0.0395 \\
Dist. from center, $r_\perp$ (core radii) & 0.1518 \\
Dist. from center, $r_\perp$ (pc) & 0.0953 \\
\hline
\end{tabular*}

\vspace{2mm}
% 底部注释部分
{\noindent \textbf{Notes:} Companion mass assumes a pulsar mass of $1.35 \, M_{\odot}$. $M_{\rm c, min}$ and $M_{\rm c, med}$ assume $i=90^\circ$ and $i=60^\circ$, respectively.}
\end{table}

%[ydj@localhost M92B]$ ./alex_latex_parfile -par "J1717+4308B_48.par" -gc_center_coord  "17:17:07.39,+43:08:09.4" -gc_distance 8.3 -gc_core_radius 0.26
%##############################################################
%#                  PSRALEX - alex_latex_parfile              #
%##############################################################
% Mp = 1.35 Msun 
%Mass Function:          3.550159e-03 Msun
%Minimum Mc:             2.047323e-01 Msun
%Median Mc:              2.399624e-01 Msun
%Distance in RA:              0.030399 arcmin
%Distance in DEC:            -0.025164 arcmin
%Total Distance:              0.039463 arcmin
%Total Distance:              0.151782 core radii
%Total Distance:              0.095279 pc (for an assumed distance of 8.300 kpc)
%##############################################################
% Weighted RMS residual: pre-fit    21.996 us. Predicted post-fit    21.996 us.
% Chisqr/nfree:    189.75/  136 =     1.395200916   pre/post:   1.00   Wmax:    6.1
%##############################################################

\begin{figure}[htp!]
\begin{center}
\includegraphics[width=1\linewidth]{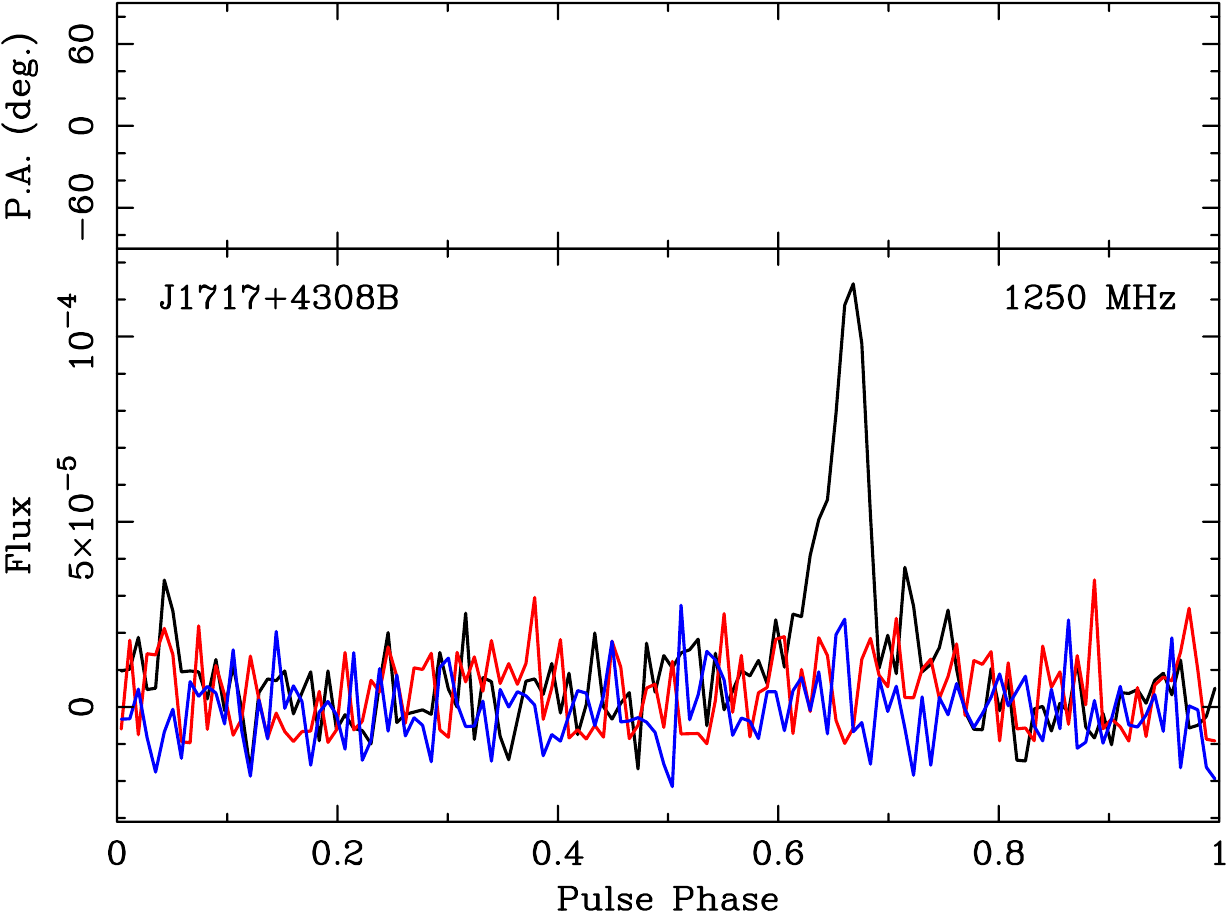}
\caption{
The polarization profile of M92B at MJD 60310.
The black, red, and blue curves correspond to the total intensity ($I$), linear ($L$), and circular ($V$) components, respectively. The upper panel presents the position angle of the linear polarization.}
\label{fig:polarization}
\end{center}
\end{figure}

Using the brightest detection of M92B at MJD 60310, we obtained the polarization profile of M92B (shown in Figure~\ref{fig:polarization}), with data folded using the \textsc{DSPSR} package \citep{2011PASA...28....1V}, cleaned with the \texttt{pazi} routine, and polarization calibrated using the \texttt{pac} routine in \textsc{PSRCHIVE} \citep{2012AR&T....9..237V}.
We searched for a rotation measure but found no significant detection.
No significant linear or circular polarization is detected at 1250\,MHz, and the polarization position angle cannot be reliably determined due to the low signal-to-noise ratio.

\section{discussion}\label{discussion} 

\subsection{Properties of new discovery}

When M92A was first discovered, its association with M92 was inferred mainly from the agreement between its DM value and the predictions of Galactic electron density models (i.e., $YMW16$; \citealt{2017ApJ...835...29Y}), together with its binary properties typical of GC pulsars \citep{2020ApJ...892L...6P}. 
A definitive confirmation was expected to require the discovery of another pulsar in the cluster with a similar DM. 
The newly discovered pulsar M92B has a DM consistent with that of M92A ($< 0.2\, \rm pc\,cm^{-3}$), lies within the cluster core radius in projection, and exhibits a negative spin period derivative consistent with acceleration in the cluster gravitational potential (see section~\ref{GC_field}).
These results strongly support that both pulsars are members of M92.

The $X$-ray sources at 0.3--8\,keV in M92 were studied in detail by \citet{2011ApJ...736..158L}. 
The closest source to M92B is CX2 at R.A. = 17:17:07.296(5), decl. = +43:08:06.95(4). Its separation from the pulsar timing position is 1$''$.23, significantly exceeding the quoted positional uncertainties\footnote{The corresponding $\chi^2 \approx 763$, calculated from the positional offset assuming independent Gaussian uncertainties in R.A. and decl.: $\chi^2 = \left(\frac{\Delta \alpha \cos\delta}{\sigma_\alpha}\right)^2 + \left(\frac{\Delta \delta}{\sigma_\delta}\right)^2$, corresponding to a $\sim \,28\,\sigma$ positional offset.}, indicating the two sources are not associated.
The derived binary parameters suggest that the companion star is a white dwarf or a low-mass star. We searched for an optical counterpart to this companion using data from the Hubble Space Telescope UV Globular Cluster Survey (HUGS; GO-10775, PI: Sarajedini; GO-13297, PI: Piotto; see \citealt{2015AJ....149...91P, 2018MNRAS.481.3382N})\footnote{\url{https://archive.stsci.edu/prepds/hugs/\#dataaccess}}. 
We crossmatched potential counterparts across five filters (F275W, F336W, F438W, F606W, and F814W) but found no significant signal within the 3\,$\sigma$ upper limits.

M92B was detected in 36 of 53 observations, and we derived its 1250 MHz flux densities using the radiometer equation \citep[e.g.,][]{1985ApJ...294L..25D} and the FAST GC pulsar survey parameters \citep[e.g.,][]{2021ApJ...915L..28P}, obtaining a range of 0.75–-4.82 $\,\mu{\rm Jy}$ with a median of 2.37 $\,\mu{\rm Jy}$. The corresponding median pseudo-luminosity, $L_{1250} \sim 0.16\, {\rm mJy\,kpc^2}$, places M92B among the faintest binary MSPs currently known in GCs. Interstellar scintillation can occasionally boost its signal above the detection threshold, allowing us to reveal this faint binary pulsar that would otherwise remain hidden.

Its faintness and intermittent detectability indicate that additional faint MSPs may remain hidden in M92.
This suggests that some pulsar-poor clusters may host substantial MSP populations dominated by sources below current detection thresholds, implying that current estimates of the GC MSP luminosity function may remain biased at the faint end (e.g., \citealt{2007ApJ...670..363H,2011MNRAS.418..477B,2013MNRAS.431..874C}).
Recent deep FAST searches using the power-spectrum stacking method have also revealed the presence of even fainter isolated MSP populations \citep{2026ApJ..1002L..31D}. 
A joint analysis of the luminosity distributions of faint binary and isolated MSPs uncovered by recent deep surveys (e.g., MeerKAT and FAST) could yield new constraints on the intrinsic GC MSP population but is beyond the scope of this Letter.

Previous studies predicted a population of $\sim 13-43$ pulsars in M92 (e.g., \citealt{2011MNRAS.418..477B, 2013MNRAS.436.3720T, 2024ApJ...969L...7Y}). 
Binary pulsars in M92 are expected to dominate the pulsar population due to the low stellar encounter rate per binary in the cluster (e.g., \citealt{2014A&A...561A..11V, 2023MNRAS.525.4167O}).
If more pulsars are present, their nondetection could be due to limited sensitivity or strong orbital acceleration in compact binary systems, which reduces the effectiveness of standard periodicity searches. 
More computationally intensive techniques, such as phase-modulation searches \citep{2003ApJ...589..911R} and template-bank acceleration searches \citep{2022MNRAS.511.1265B}, may reveal highly accelerated systems in future analyses.

The small DM of pulsars in M92 implies only modest dispersion smearing even at low radio frequencies.
Pulsars in GCs with typical steep radio spectra are expected to be brighter at low frequencies, therefore making sensitive low-frequency observations favorable for additional discoveries (e.g., \citealt{2025ApJ...988..161D, 2025arXiv251211058D}).
Discovering additional pulsars in M92 would not only improve the census of the cluster pulsar population but also provide new probes of the cluster potential, thereby further constraining its dynamical evolution (see section~\ref{mcmc}).

\subsection{Eccentricity of M92B}

M92B exhibits a mildly eccentric orbit with $e \approx 4.8 \times 10^{-4}$. 
With $P_{\rm b} \approx 2$ \, days and $M_{\rm c} \approx 0.2 \, M_{\odot}$, 
the system is consistent with the $P_{\rm b} - M_{\rm c}$ relation for helium white dwarf (He WD) companions originating from low-mass X-ray binary (LMXB) evolution \citep{1999A&A...350..928T}. 
The low eccentricity magnitude disfavors an exchange origin, which typically yields $e \gtrsim 0.1$ (e.g., \citealt{1993ApJ...415..631S}), suggesting M92B is a primordial MSP-He WD system.

The initial eccentricity in MSP progenitors is expected to dissipate via tidal circularization during the LMXB phase \citep{1992RSPTA.341...39P}. However, stellar interactions in dense GCs can subsequently induce nonzero eccentricities \citep{1995ApJ...445L.133R, 1996MNRAS.282.1064H}. Following \citet{1995ApJ...445L.133R} and \citet{2011ApJ...734...89L}, we estimate the timescale $t_{>e}$ required to induce the observed eccentricity as
\begin{equation}
\begin{split}
t_{>e} &\simeq 4\times10^{11}\ {\rm yr} \left(\frac{n}{10^4\ {\rm pc^{-3}}}\right)^{-1} \\
       & \quad \,  \times\,\left(\frac{v_0}{10\ {\rm km\ s^{-1}}}\right) \left(\frac{P_{\rm b}}{\rm days}\right)^{-2/3} e^{2/5},
\end{split}
\label{eq:te}
\end{equation}
where $n$ is the stellar number density ($n \propto \rho_0$, where $\rho_0$ is the central mass density of the GC), $v_0$ is the central stellar velocity dispersion, $P_{\rm b}$ is the orbital period, and $e$ is the observed eccentricity. 

\begin{figure}[h]
\begin{center}
\includegraphics[width=1\linewidth]{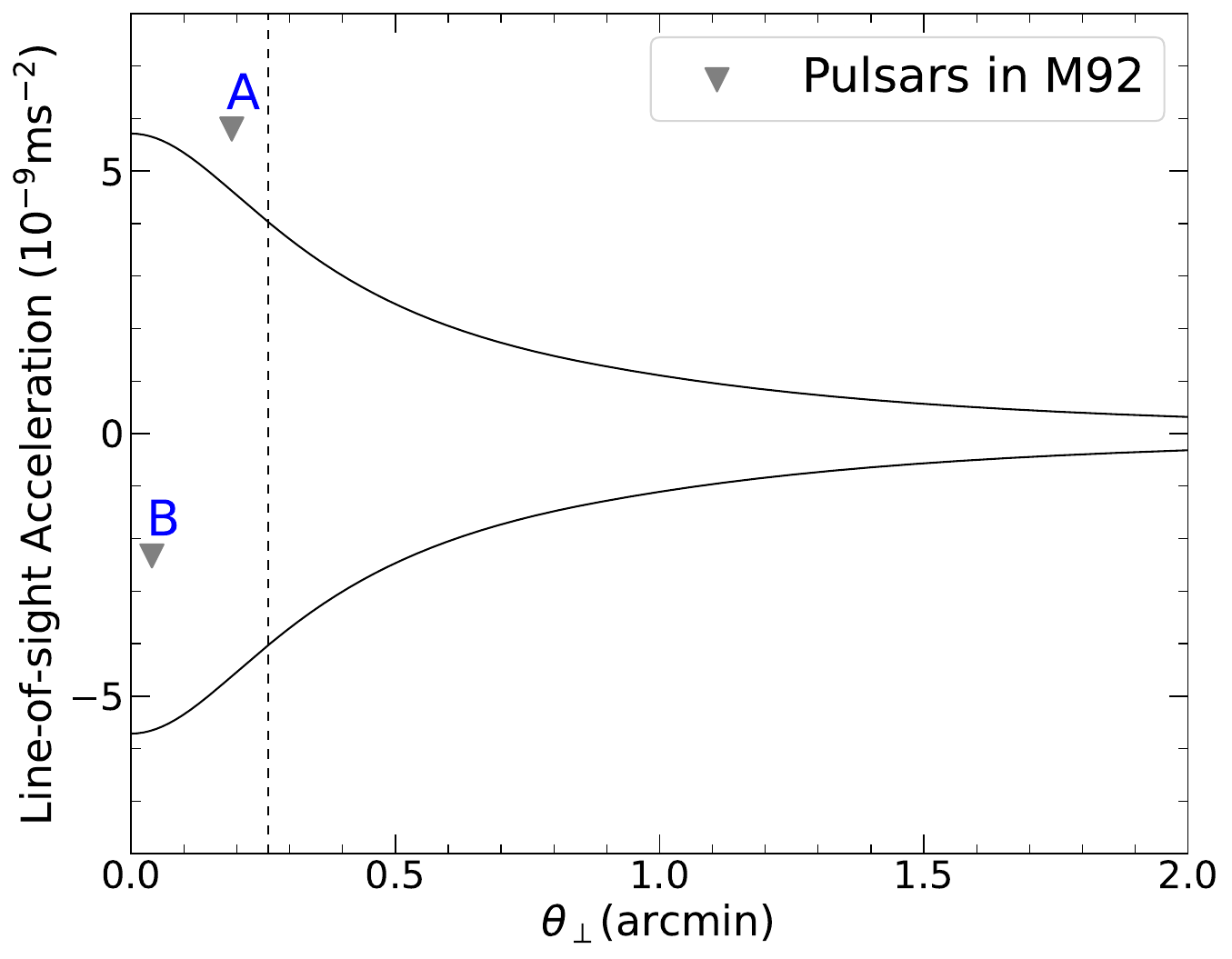}
\caption{
Acceleration model for M92. Black curves show the maximum and minimum LOS accelerations ($a_{\ell,\rm GC}$) predicted by the cluster potential as a function of projected offset ($\theta_\perp$). Downward triangles denote independent limits on pulsar accelerations ($a_{\ell,P,\max}$), and the vertical dashed line shows the core radius of M92 ($r_{\rm c}=0.26'$). The value of M92A is from \citet{2025ApJS..279...51L}. 
%The vertical dashed line shows the core radius of M92 ($r_{\rm c}=0.26'$).
}
\label{fig:M92_Acceleration}
\end{center}
\end{figure}

As a first approximation, we adopt the central mass density $\rho_{\rm c}$ and convert it to a number density via
$
n \approx {\rho_{\rm 0}} /{\langle m_\star \rangle},
$
where $\langle m_\star \rangle$ is the mean stellar mass, taken to be $\sim 1\,M_\odot$ following \citet{2011ApJ...734...89L}. 
For M92B, adopting $\rho_0 \approx 1.995 \times 10^4\,L_{\odot}\,{\rm pc^{-3}}$ \citep{2010arXiv1012.3224H} and $v_0 \approx 8.4\,{\rm km\ s^{-1}}$ \citep{2022ApJ...934..150L}, we estimate $t_{>e} \approx 4.55\,{\rm Gyr}$.
This timescale is shorter than the cluster age of $12.75\pm0.25$\,Gyr \citep{2007AJ....133.2787P}, indicating that dynamical perturbations are sufficient to produce the observed eccentricity.

\subsection{Acceleration of M92B Caused by the Cluster Potential}\label{GC_field}

The observed spin period derivative of M92B is negative ($\dot{P}_{\rm obs}<0$). Since radio pulsars intrinsically spin down, a negative $\dot{P}_{\rm obs}$ implies that the gravitational potential of the cluster dominates the LOS acceleration of M92B (e.g., \citealt{1993ASPC...50..141P}). For pulsars in GCs, the observed period derivative can be written as
\begin{equation}
\left( \frac{\dot{P}}{P} \right)_{\rm obs}
=
\left( \frac{\dot{P}}{P} \right)_{\rm int}
+
\frac{\mu^2 d}{c}
+
\frac{a_{\ell,\,\rm GC}}{c}
+
\frac{a_g}{c},
\end{equation}
where $\mu$ is the total proper motion, $d$ is the cluster distance (8.3\,kpc; \citealt{2010arXiv1012.3224H}), and $\mu^2 d/c$ represents the Shklovskii effect \citep{1970SvA....13..562S}. The terms $a_{\ell,\rm GC}$ and $a_g$ denote the LOS acceleration due to the cluster potential and the differential Galactic acceleration between the solar system and the GC, respectively.

The proper motion of M92B has not yet been measured, but it is expected to be similar to that of M92. Using the $Gaia$ EDR3 measurement of the cluster proper motion ($-4.935 \pm 0.024$ and $-0.625 \pm 0.024\,\rm mas\,yr^{-1}$; \citealt{2021MNRAS.505.5978V}), the Shklovskii contribution is estimated to be $\mu^2 d \approx 1.50 \times 10^{-10}\,\rm m\,s^{-2}$.

The Galactic acceleration term can be estimated following \citet{1995ApJ...441..429N} and \citet{2017ApJ...845..148P}:
\begin{eqnarray}
a_{g}\cdot\vec{n} =
-{\rm cos}(b)
\left(\frac{\Theta_0^2}{R_0}\right)
\left(
{\rm cos}(l)+
\frac{\beta}{{\rm sin}^2(l)+\beta^2}
\right),
\end{eqnarray}
where $R_0=8.34 \, \pm 0.16 \, $ \,kpc and $\Theta_0=240 \,\pm8 \,\rm \, km\,s^{-1}$ are the solar Galactocentric distance and circular velocity \citep{2014ApJ...793...51S}. Here $\beta=(d/R_0)\cos b-\cos l$, and $l$ and $b$ are the Galactic longitude and latitude. For M92 ($l=68.34^\circ$, $b=34.86^\circ$), this yields $a_g \approx -1.45\times10^{-10}\,\rm m\,s^{-2}$.

To estimate the cluster acceleration $a_{\ell,\rm GC}$, we adopt the analytical model derived from \citet{2005ApJ...621..959F}, which assumes the spatial \citet{1962AJ.....67..471K} density profile:
\begin{equation}
 a_{\ell,\,\rm GC}(x)
 =
 \frac{9v_{\rm 0}^2}{d\theta_c}
 \frac{\ell}{x^3}
 \left(
 \frac{x}{\sqrt{1+x^2}}-\sinh^{-1}x
 \right),
\end{equation}
where $x$ is the distance from the pulsar to the center of the GC divided by its core radius ($r_c=\theta_c d$), and $\ell$ is the LOS offset from the cluster center, also in units of $r_c$. For each projected offset $\theta_\perp$, the model provides the maximum allowed LOS acceleration $a_{\ell,\rm max}$ (shown in Figure~\ref{fig:M92_Acceleration}).

An independent constraint can be derived from the observed period derivative by assuming $\dot{P}_{\rm int}=0$:
\begin{equation}
a_{\ell,\rm P,max}
=
c\frac{\dot{P}_{\rm obs}}{P}
-
\mu^2 d
-
a_g .
\end{equation}
For M92B we obtain $a_{\ell,\rm P,max}\approx -5.65\times10^{-9}\,\rm m\,s^{-2}$.
The Shklovskii term and Galactic acceleration are orders of magnitude smaller than the cluster potential and the observed $c\frac{\dot{P}_{\rm obs}}{P}$ of M92B, and thus can be neglected. This suggests that the cluster gravitational potential model can well account for its negative $\dot{P}_{\rm obs}$.

Assuming the pulsar acceleration lies within $\pm a_{\ell,\rm max}$, one can derive the limits for $\dot{P}_{\rm int}$. For M92B, we obtain $\dot{P}_{\rm int}< 3.9 \times10^{-20}\,\rm s\,s^{-1}$, corresponding to a surface magnetic field $B_s = 3.2\times 10^{19}(P\dot{P})^{1/2}\ G < 3.7\times10^8\,\rm G$ and a characteristic age $\tau_c = \frac{P}{2\dot{P}} > 1.4\,\rm Gyr$. 
Given that the M92's age is $12.75\pm0.25$\,Gyr \citep{2007AJ....133.2787P}, the significant $\tau_c$ of  
M92B, compared to that of M92A ($\tau_c<1.36$~Gyr; \citealt{2025ApJS..279...51L}), implies that its real age is not constrained.
Thus, we cannot exclude that M92B formed during the early stages of the cluster's evolution.
This is also roughly consistent with the time scale $t_{>e} \approx 4.55\,{\rm Gyr}$ for M92B, further indicating that this system is likely old.
For binaries, orbital period derivative measurements can provide tighter constraints on cluster acceleration and intrinsic spin-down \citep[e.g.,][]{2017MNRAS.471..857F}, although no significant detection is obtained with the current timing baseline. Preliminary \textsc{PINT} simulations \citep{2021ApJ...911...45L} suggest that a significant detection of the expected acceleration-induced $\dot{P}_{\rm b}\sim -10^{-12}$ from the cluster gravitational potential would likely require roughly another decade of timing observations.

\begin{figure}[htp!]
%\begin{center}
\includegraphics[width=1\linewidth]{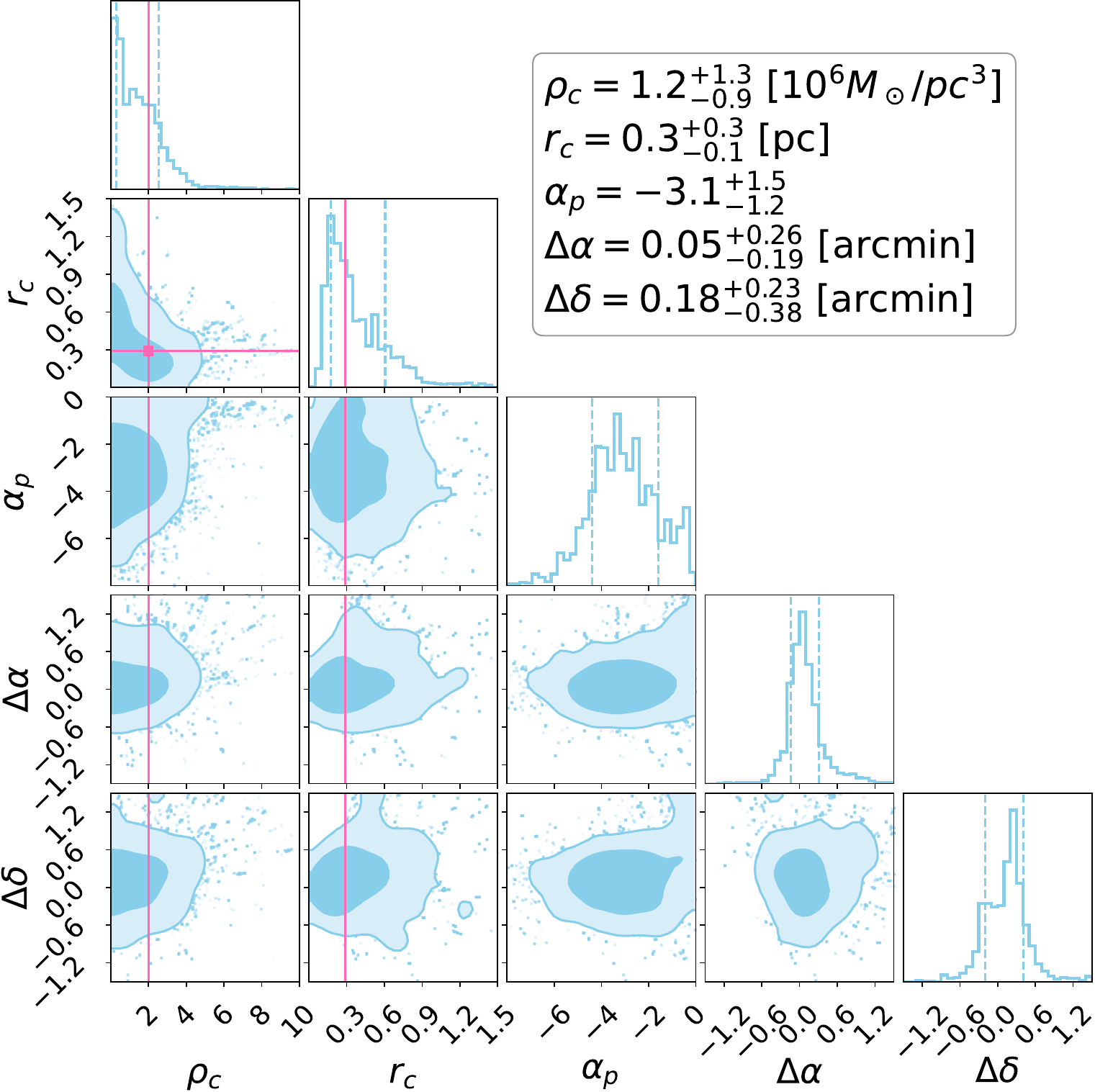}
\caption{Marginalized posterior probability distributions for the M92 structural parameters and the offset from the gravitational center of M92 \citep{2021MNRAS.505.5978V}. Parameters shown include the central mass density ($\rho_c$ in $10^6\,M_\odot\,\rm pc^{-3}$), core radius ($r_c$ in pc), mass-segregation index of pulsars ($\alpha_p$), and positional offsets from the \textit{Gaia}-measured center ($\Delta \alpha$, $\Delta \delta$ in arcmin). The values shown in the legend are posterior medians with \(1\sigma\) uncertainties, defined by the 16th--84th percentile range.
In the 2D panels, the contours represent the $1\sigma$ and $2\sigma$ confidence level. The vertical dashed lines in the one-dimensional histograms mark the 16th and 84th percentiles, while the solid pink lines and markers denote the cluster core parameters derived from $N$-body dynamical modeling \citep{baumgardt2018catalogue} for comparison.}

\label{fig:constraints}
%\end{center}
\end{figure}

\subsection{MCMC Constraint of the Cluster dynamics}\label{mcmc}

To resolve degeneracies in photometric core parameters and provide an independent dynamical measurement of M92 from pulsars, we performed a Bayesian Markov Chain Monte Carlo (MCMC) analysis based on the framework of \citet{2017ApJ...845..148P} and \citet{2018MNRAS.481..627A,2019ApJ...884L...9A}. We model mass segregation via a generalized pulsar number density, $n_p(r) \propto [1+(r/r_c)^2]^{\alpha_p}$, implementing the likelihood and density modeling formalisms defined in Equations (30)--(38) of \citet{2017ApJ...845..148P}.

The MCMC model contains seven free parameters: $\rho_c,\ r_c,\ \alpha_p,\ \Delta\alpha,\ \Delta\delta,\ \ell_{\rm A}$, and $\ell_{\rm B}$,
where \(\rho_c\) and \(r_c\) are the central mass density and core radius, \(\alpha_p\) is the pulsar mass-segregation index, \(\Delta\alpha\) and \(\Delta\delta\) are the offsets of the trial gravitational center from the Gaia EDR3, and \(\ell_{\rm A}\) and \(\ell_{\rm B}\) are the unknown LOS positions of M92A and M92B.
For each pulsar, the likelihood uses its measured sky position and LOS acceleration inferred from the observed $(\dot{P}/P)_{\rm obs}$, after correcting for the Shklovskii term, Galactic acceleration, and a fiducial intrinsic spin-down contribution. The latter is estimated from a magnetic-dipole model with the median field-MSP magnetic field fixed at $B=10^{8.47}\,\mathrm{G}$ \citep{2017ApJ...845..148P,2018MNRAS.481..627A,2019ApJ...884L...9A}. The sky positions determine the projected offsets from each trial cluster center and enter the spatial-density weighting, whereas the period derivatives constrain the LOS gravitational acceleration. The unknown LOS positions, \(\ell_{\rm A}\) and \(\ell_{\rm B}\), are sampled and marginalized over.

The resulting posterior distributions (see Figure~\ref{fig:constraints}, corresponding to the best-fit values with 1$\sigma$ errors) yield a central mass density of $\rho_c = 1.2_{-0.9}^{+1.3} \times 10^6 \, M_\odot\,\text{pc}^{-3}$ and a core radius of $r_c = 0.3_{-0.1}^{+0.3}$~pc. These values are in good agreement with the structural parameters derived from $N$-body dynamical modeling ($\rho_c = 2 \times 10^6 \, M_\odot\,\text{pc}^{-3}$ and $r_c = 0.29$~pc; \citealt{baumgardt2018catalogue}) within 1$\sigma$.
The inferred center-of-gravity offsets are, $\Delta\alpha = 0.05_{-0.19}^{+0.26}$~arcmin and $\Delta\delta = 0.18_{-0.38}^{+0.23}$~arcmin. Both offsets are consistent with zero within $1\sigma$, indicating that the current two-pulsar sample does not provide a statistically significant constraint on the cluster gravitational center. The pulsar mass-segregation index is $\alpha_p = -3.1_{-1.2}^{+1.5}$, consistent with the constraint of $\alpha_p = -3.56_{-0.56}^{+0.47}$ derived from 36 pulsars in Terzan 5 \citep{2017ApJ...845..148P}. These findings highlight the utility of pulsars as independent dynamical probes that bypass the biases of photometric measurements. As the pulsar sample in M92 grows, these fundamental structural parameters are expected to be constrained with higher precision.

\section{Conclusions}\label{Conclusions} 

In this work, we conducted a search for additional pulsars in the GC M92 using 6\,yr years of FAST timing observations originally obtained for M92A. 
This 6\,yr observational baseline significantly increases the possibility of discovering faint pulsars, particularly those intermittently detectable due to interstellar scintillation. Our main results are summarized as follows:

\begin{enumerate}
\item We discovered a new MSP in M92, PSR~J1717+4308B (M92B). This pulsar has a spin period of 3.51\,ms and resides in a binary system with an orbital period of 2.3\,days and a mildly eccentric orbit ($e \simeq 4.8 \times 10^{-4}$). The minimum companion mass is $M_c\approx 0.2\,M_\odot$, suggesting a He WD companion.

\item The DM value of M92B is consistent with that of M92A ($< 0.2\,\mathrm{pc\,cm^{-3}}$), its projected position lies within the cluster core radius, and its negative spin period derivative is consistent with acceleration in the cluster potential, all supporting their cluster membership.

\item No $X$-ray, optical, or UV band counterpart to M92B is identified. The nearest $X$-ray source (CX2) is positionally inconsistent, and no counterpart is detected in archival Hubble Space Telescope data down to 3\,$\sigma$ upper limits.

\item A Bayesian MCMC analysis using M92A and M92B yields broad constraints on the core structural parameters of M92, broadly consistent with \textit{N}-body dynamical modeling. This demonstrates the utility of pulsar timing for providing dynamical constraints even in clusters with very limited pulsar samples.

\item Given previous estimates of 13–-43 pulsars in M92, additional pulsars likely remain undetected. If such a population indeed exists, their nondetection suggests these undiscovered sources are intrinsically faint, implying incompleteness in the faint end of the GC MSP luminosity distribution. More sensitive observations and advanced search techniques are expected to reveal further pulsars, thereby improving constraints on cluster dynamics.

\end{enumerate}

\begin{acknowledgements}
This work is supported by Guizhou provincial natural science foundation project No. ZD[2026]058.
Liyun Zhang has been supported by the Science and Technology Program of Guizhou Province under project Nos. QKHPTRC-ZDSYS[2023]003 and QKHFQ[2023]003.
This work is also supported by the Guizhou Provincial Major Scientific and Technological Program (Nos. XKBF (2025)010 and XKBF (2025)011).
This work made use of the data from FAST (Five-hundred-meter Aperture Spherical radio Telescope) (https://cstr.cn/31116.02.FAST).  FAST is a Chinese national mega-science facility, operated by National Astronomical Observatories, Chinese Academy of Sciences.
We gratefully acknowledge the generous computational support provided by Guizhou Suanjia Computing Services Co., Ltd, Gui'an New Area Science and Technology Innovation Industries Development Limited Company, and Gui'an Supercomputing Center.
We are grateful to Dr. Zhichen Pan and Dr. Lei Qian for their guidance on pulsar searching and timing analyses, to Tong Liu and Ruili He for discussions on polarization calibration, and to Yujie Chen for help with timing simulation.
Finally, we thank the anonymous referee for many constructive comments and valuable suggestions that greatly improved the clarity and quality of the manuscript.
\end{acknowledgements}

% Facility: FAST. Software:PRESTO (Ransom et al. 2002), DSPSR (van Straten \& Bailes 2011), TEMPO2 (Hobbs et al. 2006).

% \vspace{5mm}
% \facilities{HST(STIS), Swift(XRT and UVOT), AAVSO, CTIO:1.3m, CTIO:1.5m, CXO}


% \software{astropy \citep{2013A&A...558A..33A,2018AJ....156..123A},  
%          Cloudy \citep{2013RMxAA..49..137F}, 
%          Source Extractor \citep{1996A&AS..117..393B}
%          }



\bibliography{Refs_M92B}{}

@article{baumgardt2018catalogue,
  title={A catalogue of masses, structural parameters, and velocity dispersion profiles of 112 Milky Way globular clusters},
  author={Baumgardt, Holger and Hilker, Michael},
  journal={Monthly Notices of the Royal Astronomical Society},
  volume={478},
  number={2},
  pages={1520--1557},
  year={2018},
  publisher={Oxford University Press}
}

@ARTICLE{2019ApJ...884L...9A,
       author = {{Abbate}, Federico and {Possenti}, Andrea and {Colpi}, Monica and {Spera}, Mario},
        title = "{Evidence of Nonluminous Matter in the Center of M62}",
      journal = {\apjl},
         year = 2019,
        month = oct,
       volume = {884},
       number = {1},
          eid = {L9},
        pages = {L9},
          doi = {10.3847/2041-8213/ab46c3},
archivePrefix = {arXiv},
       eprint = {1909.11091},
 primaryClass = {astro-ph.HE},
       adsurl = {https://ui.adsabs.harvard.edu/abs/2019ApJ...884L...9A}
}

@ARTICLE{2000ApJ...535..975C,
       author = {{Camilo}, F. and {Lorimer}, D.~R. and {Freire}, P. and {Lyne}, A.~G. and {Manchester}, R.~N.},
        title = "{Observations of 20 Millisecond Pulsars in 47 Tucanae at 20 Centimeters}",
      journal = {\apj},
         year = 2000,
        month = jun,
       volume = {535},
       number = {2},
        pages = {975-990},
          doi = {10.1086/308859},
archivePrefix = {arXiv},
       eprint = {astro-ph/9911234},
 primaryClass = {astro-ph},
       adsurl = {https://ui.adsabs.harvard.edu/abs/2000ApJ...535..975C}
}

@ARTICLE{2026ApJ..1002L..31D,
       author = {{Dai}, Yinfeng and {Zhu}, Xing-Jiang and {Pan}, Zhichen and {Qian}, Lei and {Zhang}, Li-yun and {Yin}, Dejiang and {Pan}, Yu and {Peng}, Bo and {Li}, Baoda and {Lian}, Yujie and {Li}, Yaowei and {Wu}, Yuxiao and {Huang}, Menglin and {Hao}, Qiaoli and {Wang}, Xingyi and {Niu}, Xianghua and {Song}, Jinyou and {Guo}, Minglei and {Chen}, Shuangyuan},
        title = "{The Stack Search Tests on FAST Data: Discovery of Six Faint Isolated Millisecond Pulsars in NGC 6517 and NGC 7078 (M15)}",
      journal = {\apjl},
         year = 2026,
        month = may,
       volume = {1002},
       number = {1},
          eid = {L31},
        pages = {L31},
          doi = {10.3847/2041-8213/ae5dbb},
archivePrefix = {arXiv},
       eprint = {2604.08268},
 primaryClass = {astro-ph.HE},
       adsurl = {https://ui.adsabs.harvard.edu/abs/2026ApJ..1002L..31D}
}

@ARTICLE{2015AJ....149...91P,
       author = {{Piotto}, G. and {Milone}, A.~P. and {Bedin}, L.~R. and {Anderson}, J. and {King}, I.~R. and {Marino}, A.~F. and {Nardiello}, D. and {Aparicio}, A. and {Barbuy}, B. and {Bellini}, A. and {Brown}, T.~M. and {Cassisi}, S. and {Cool}, A.~M. and {Cunial}, A. and {Dalessandro}, E. and {D'Antona}, F. and {Ferraro}, F.~R. and {Hidalgo}, S. and {Lanzoni}, B. and {Monelli}, M. and {Ortolani}, S. and {Renzini}, A. and {Salaris}, M. and {Sarajedini}, A. and {van der Marel}, R.~P. and {Vesperini}, E. and {Zoccali}, M.},
        title = "{The Hubble Space Telescope UV Legacy Survey of Galactic Globular Clusters. I. Overview of the Project and Detection of Multiple Stellar Populations}",
      journal = {\aj},
         year = 2015,
        month = mar,
       volume = {149},
       number = {3},
          eid = {91},
        pages = {91},
          doi = {10.1088/0004-6256/149/3/91},
archivePrefix = {arXiv},
       eprint = {1410.4564},
 primaryClass = {astro-ph.SR},
       adsurl = {https://ui.adsabs.harvard.edu/abs/2015AJ....149...91P}
}

@ARTICLE{2018MNRAS.481.3382N,
       author = {{Nardiello}, D. and {Libralato}, M. and {Piotto}, G. and {Anderson}, J. and {Bellini}, A. and {Aparicio}, A. and {Bedin}, L.~R. and {Cassisi}, S. and {Granata}, V. and {King}, I.~R. and {Lucertini}, F. and {Marino}, A.~F. and {Milone}, A.~P. and {Ortolani}, S. and {Platais}, I. and {van der Marel}, R.~P.},
        title = "{The Hubble Space Telescope UV Legacy Survey of Galactic Globular Clusters - XVII. Public Catalogue Release}",
      journal = {\mnras},
         year = 2018,
        month = dec,
       volume = {481},
       number = {3},
        pages = {3382-3393},
          doi = {10.1093/mnras/sty2515},
archivePrefix = {arXiv},
       eprint = {1809.04300},
 primaryClass = {astro-ph.SR},
       adsurl = {https://ui.adsabs.harvard.edu/abs/2018MNRAS.481.3382N}
}

@ARTICLE{2011ApJ...736..158L,
       author = {{Lu}, Ting-Ni and {Kong}, Albert K.~H. and {Verbunt}, Frank and {Lewin}, Walter H.~G. and {Anderson}, Scott F. and {Pooley}, David},
        title = "{Chandra and HST Studies of the X-Ray Sources in Galactic Globular Cluster M92}",
      journal = {\apj},
         year = 2011,
        month = aug,
       volume = {736},
       number = {2},
          eid = {158},
        pages = {158},
          doi = {10.1088/0004-637X/736/2/158},
       adsurl = {https://ui.adsabs.harvard.edu/abs/2011ApJ...736..158L}
}

@ARTICLE{2016RAA....16..151Z,
       author = {{Zhang}, Lei and {Hobbs}, George and {Li}, Di and {Lorimer}, Duncan and {Zhang}, Jie and {Yu}, Meng and {Yue}, You-Ling and {Wang}, Pei and {Pan}, Zhi-Chen and {Dai}, Shi},
        title = "{Wide-bandwidth drift-scan pulsar surveys of globular clusters: application to early science observations with FAST}",
      journal = {Research in Astronomy and Astrophysics},
         year = 2016,
        month = oct,
       volume = {16},
       number = {10},
          eid = {151},
        pages = {151},
          doi = {10.1088/1674-4527/16/10/151},
       adsurl = {https://ui.adsabs.harvard.edu/abs/2016RAA....16..151Z}
}

@ARTICLE{2025ApJS..279...51L,
       author = {{Lian}, Yujie and {Pan}, Zhichen and {Zhang}, Haiyan and {Cao}, Shuo and {Freire}, P.~C.~C. and {Qian}, Lei and {Eatough}, Ralph P. and {Shao}, Lijing and {Ransom}, Scott M. and {Lorimer}, Duncan R. and {Yin}, Dejiang and {Dai}, Yinfeng and {Liu}, Kuo and {Wang}, Lin and {Wang}, Yujie and {Zhang}, Zhongli and {Feng}, Zhonghua and {Li}, Baoda and {Li}, Minghui and {Liu}, Tong and {Li}, Yaowei and {Peng}, Bo and {Pan}, Yu and {Wu}, Yuxiao and {Zhang}, Liyun and {Zhang}, Xingnan and {Jiang}, Peng},
        title = "{The FAST Globular Cluster Pulsar Survey (GC FANS)}",
      journal = {\apjs},
         year = 2025,
        month = aug,
       volume = {279},
       number = {2},
          eid = {51},
        pages = {51},
          doi = {10.3847/1538-4365/ade4ba},
archivePrefix = {arXiv},
       eprint = {2506.07970},
 primaryClass = {astro-ph.HE},
       adsurl = {https://ui.adsabs.harvard.edu/abs/2025ApJS..279...51L}
}

@ARTICLE{2006Sci...311.1901H,
       author = {{Hessels}, Jason W.~T. and {Ransom}, Scott M. and {Stairs}, Ingrid H. and {Freire}, Paulo C.~C. and {Kaspi}, Victoria M. and {Camilo}, Fernando},
        title = "{A Radio Pulsar Spinning at 716 Hz}",
      journal = {Science},
         year = 2006,
        month = mar,
       volume = {311},
       number = {5769},
        pages = {1901-1904},
          doi = {10.1126/science.1123430},
archivePrefix = {arXiv},
       eprint = {astro-ph/0601337},
 primaryClass = {astro-ph},
       adsurl = {https://ui.adsabs.harvard.edu/abs/2006Sci...311.1901H}
}

@ARTICLE{1975ApJ...199L.143C,
       author = {{Clark}, G.~W.},
        title = "{X-ray binaries in globular clusters.}",
      journal = {\apjl},
         year = 1975,
        month = aug,
       volume = {199},
        pages = {L143-L145},
          doi = {10.1086/181869},
       adsurl = {https://ui.adsabs.harvard.edu/abs/1975ApJ...199L.143C}
}

@ARTICLE{2022MNRAS.511.1265B,
       author = {{Balakrishnan}, Vishnu and {Champion}, David and {Barr}, Ewan and {Kramer}, Michael and {Venkatraman Krishnan}, V. and {Eatough}, Ralph P. and {Sengar}, Rahul and {Bailes}, Matthew},
        title = "{Coherent search for binary pulsars across all Five Keplerian parameters in radio observations using the template-bank algorithm}",
      journal = {\mnras},
         year = 2022,
        month = mar,
       volume = {511},
       number = {1},
        pages = {1265-1284},
          doi = {10.1093/mnras/stab3746},
archivePrefix = {arXiv},
       eprint = {2112.11991},
 primaryClass = {astro-ph.IM},
       adsurl = {https://ui.adsabs.harvard.edu/abs/2022MNRAS.511.1265B}
}

@ARTICLE{2003ApJ...589..911R,
       author = {{Ransom}, Scott M. and {Cordes}, James M. and {Eikenberry}, Stephen S.},
        title = "{A New Search Technique for Short Orbital Period Binary Pulsars}",
      journal = {\apj},
         year = 2003,
        month = jun,
       volume = {589},
       number = {2},
        pages = {911-920},
          doi = {10.1086/374806},
archivePrefix = {arXiv},
       eprint = {astro-ph/0210010},
 primaryClass = {astro-ph},
       adsurl = {https://ui.adsabs.harvard.edu/abs/2003ApJ...589..911R}
}

@ARTICLE{1995ApJ...441..429N,
       author = {{Nice}, D.~J. and {Taylor}, J.~H.},
        title = "{PSR J2019+2425 and PSR J2322+2057 and the Proper Motions of Millisecond Pulsars}",
      journal = {\apj},
         year = 1995,
        month = mar,
       volume = {441},
        pages = {429},
          doi = {10.1086/175367},
       adsurl = {https://ui.adsabs.harvard.edu/abs/1995ApJ...441..429N}
}

@ARTICLE{2017ApJ...845..148P,
       author = {{Prager}, Brian J. and {Ransom}, Scott M. and {Freire}, Paulo C.~C. and {Hessels}, Jason W.~T. and {Stairs}, Ingrid H. and {Arras}, Phil and {Cadelano}, Mario},
        title = "{Using Long-term Millisecond Pulsar Timing to Obtain Physical Characteristics of the Bulge Globular Cluster Terzan 5}",
      journal = {\apj},
         year = 2017,
        month = aug,
       volume = {845},
       number = {2},
          eid = {148},
        pages = {148},
          doi = {10.3847/1538-4357/aa7ed7},
archivePrefix = {arXiv},
       eprint = {1612.04395},
 primaryClass = {astro-ph.SR},
       adsurl = {https://ui.adsabs.harvard.edu/abs/2017ApJ...845..148P}
}

@ARTICLE{2014ApJ...793...51S,
       author = {{Sharma}, S. and {Bland-Hawthorn}, J. and {Binney}, J. and {Freeman}, K.~C. and {Steinmetz}, M. and {Boeche}, C. and {Bienaym{\'e}}, O. and {Gibson}, B.~K. and {Gilmore}, G.~F. and {Grebel}, E.~K. and {Helmi}, A. and {Kordopatis}, G. and {Munari}, U. and {Navarro}, J.~F. and {Parker}, Q.~A. and {Reid}, W.~A. and {Seabroke}, G.~M. and {Siebert}, A. and {Watson}, F. and {Williams}, M.~E.~K. and {Wyse}, R.~F.~G. and {Zwitter}, T.},
        title = "{Kinematic Modeling of the Milky Way Using the RAVE and GCS Stellar Surveys}",
      journal = {\apj},
         year = 2014,
        month = sep,
       volume = {793},
       number = {1},
          eid = {51},
        pages = {51},
          doi = {10.1088/0004-637X/793/1/51},
archivePrefix = {arXiv},
       eprint = {1405.7435},
 primaryClass = {astro-ph.GA},
       adsurl = {https://ui.adsabs.harvard.edu/abs/2014ApJ...793...51S}
}

@ARTICLE{2022ApJ...934..150L,
       author = {{Libralato}, Mattia and {Bellini}, Andrea and {Vesperini}, Enrico and {Piotto}, Giampaolo and {Milone}, Antonino P. and {van der Marel}, Roeland P. and {Anderson}, Jay and {Aparicio}, Antonio and {Barbuy}, Beatriz and {Bedin}, Luigi R. and {Borsato}, Luca and {Cassisi}, Santi and {Dalessandro}, Emanuele and {Ferraro}, Francesco R. and {King}, Ivan R. and {Lanzoni}, Barbara and {Nardiello}, Domenico and {Ortolani}, Sergio and {Sarajedini}, Ata and {Sohn}, Sangmo Tony},
        title = "{The Hubble Space Telescope UV Legacy Survey of Galactic Globular Clusters. XXIII. Proper-motion Catalogs and Internal Kinematics}",
      journal = {\apj},
         year = 2022,
        month = aug,
       volume = {934},
       number = {2},
          eid = {150},
        pages = {150},
          doi = {10.3847/1538-4357/ac7727},
archivePrefix = {arXiv},
       eprint = {2206.09924},
 primaryClass = {astro-ph.GA},
       adsurl = {https://ui.adsabs.harvard.edu/abs/2022ApJ...934..150L}
}

@ARTICLE{2021MNRAS.505.5978V,
       author = {{Vasiliev}, Eugene and {Baumgardt}, Holger},
        title = "{Gaia EDR3 view on galactic globular clusters}",
      journal = {\mnras},
         year = 2021,
        month = aug,
       volume = {505},
       number = {4},
        pages = {5978-6002},
          doi = {10.1093/mnras/stab1475},
archivePrefix = {arXiv},
       eprint = {2102.09568},
 primaryClass = {astro-ph.GA},
       adsurl = {https://ui.adsabs.harvard.edu/abs/2021MNRAS.505.5978V}
}

@ARTICLE{1970SvA....13..562S,
       author = {{Shklovskii}, I.~S.},
        title = "{Possible Causes of the Secular Increase in Pulsar Periods.}",
      journal = {\sovast},
         year = 1970,
        month = feb,
       volume = {13},
        pages = {562},
       adsurl = {https://ui.adsabs.harvard.edu/abs/1970SvA....13..562S}
}

@ARTICLE{2024ApJ...969L...7Y,
       author = {{Yin}, Dejiang and {Zhang}, Li-yun and {Qian}, Lei and {Eatough}, Ralph P. and {Li}, Baoda and {Lorimer}, Duncan R. and {Dai}, Yinfeng and {Li}, Yaowei and {Zhang}, Xingnan and {Li}, Minghui and {Su}, Tianhao and {Wu}, Yuxiao and {Pan}, Yu and {Lian}, Yujie and {Liu}, Tong and {Yan}, Zhen and {Pan}, Zhichen},
        title = "{FAST Discovery of Eight Isolated Millisecond Pulsars in NGC 6517}",
      journal = {\apjl},
         year = 2024,
        month = jul,
       volume = {969},
       number = {1},
          eid = {L7},
        pages = {L7},
          doi = {10.3847/2041-8213/ad534e},
archivePrefix = {arXiv},
       eprint = {2405.18228},
 primaryClass = {astro-ph.HE},
       adsurl = {https://ui.adsabs.harvard.edu/abs/2024ApJ...969L...7Y}
}

@BOOK{2004hpa..book.....L,
       author = {{Lorimer}, D.~R. and {Kramer}, M.},
        title = "{Handbook of Pulsar Astronomy}",
         year = 2004,
       volume = {4},
       adsurl = {https://ui.adsabs.harvard.edu/abs/2004hpa..book.....L}
}

@ARTICLE{2025ApJ...991...38L,
       author = {{Li}, Yaowei and {Wang}, Lin and {Qian}, Lei and {Zhang}, Liyun and {Chen}, Yujie and {Yin}, Dejiang and {Li}, Baoda and {Dai}, Yinfeng and {Eatough}, Ralph P. and {Li}, Wenze and {Jiang}, Dongyue and {Zhang}, Xingnan and {Li}, Minghui and {Lian}, Yujie and {Wu}, Yuxiao and {Liu}, Tong and {Liu}, Kuo and {Pan}, Zhichen},
        title = "{Searching for Pulsars in Globular Clusters with the Fast-folding Algorithm and a New Pulsar Discovered in M13}",
      journal = {\apj},
         year = 2025,
        month = sep,
       volume = {991},
       number = {1},
          eid = {38},
        pages = {38},
          doi = {10.3847/1538-4357/add6a5},
archivePrefix = {arXiv},
       eprint = {2505.05021},
 primaryClass = {astro-ph.HE},
       adsurl = {https://ui.adsabs.harvard.edu/abs/2025ApJ...991...38L}
}

@ARTICLE{2025ApJ...991..177Y,
       author = {{Yin}, Dejiang and {Wang}, Lin and {Zhang}, Li-yun and {Qian}, Lei and {Li}, Baoda and {Liu}, Kuo and {Peng}, Bo and {Dai}, Yinfeng and {Li}, Yaowei and {Pan}, Zhichen},
        title = "{Illuminating Hidden Pulsars: Scintillation-enhanced Discovery of Two Binary Millisecond Pulsars in M13 with FAST}",
      journal = {\apj},
         year = 2025,
        month = oct,
       volume = {991},
       number = {2},
          eid = {177},
        pages = {177},
          doi = {10.3847/1538-4357/adfa14},
archivePrefix = {arXiv},
       eprint = {2508.05998},
 primaryClass = {astro-ph.HE},
       adsurl = {https://ui.adsabs.harvard.edu/abs/2025ApJ...991..177Y}
}

@ARTICLE{2018MNRAS.481..627A,
       author = {{Abbate}, F. and {Possenti}, A. and {Ridolfi}, A. and {Freire}, P.~C.~C. and {Camilo}, F. and {Manchester}, R.~N. and {D'Amico}, N.},
        title = "{Internal gas models and central black hole in 47 Tucanae using millisecond pulsars}",
      journal = {\mnras},
         year = 2018,
        month = nov,
       volume = {481},
       number = {1},
        pages = {627-638},
          doi = {10.1093/mnras/sty2298},
archivePrefix = {arXiv},
       eprint = {1808.06621},
 primaryClass = {astro-ph.HE},
       adsurl = {https://ui.adsabs.harvard.edu/abs/2018MNRAS.481..627A}
}

@ARTICLE{2001ApJ...557L.105F,
       author = {{Freire}, P.~C. and {Kramer}, M. and {Lyne}, A.~G. and {Camilo}, F. and {Manchester}, R.~N. and {D'Amico}, N.},
        title = "{Detection of Ionized Gas in the Globular Cluster 47 Tucanae}",
      journal = {\apjl},
         year = 2001,
        month = aug,
       volume = {557},
       number = {2},
        pages = {L105-L108},
          doi = {10.1086/323248},
archivePrefix = {arXiv},
       eprint = {astro-ph/0107206},
 primaryClass = {astro-ph},
       adsurl = {https://ui.adsabs.harvard.edu/abs/2001ApJ...557L.105F}
}

@ARTICLE{2024ApJ...972..198C,
       author = {{Corongiu}, A. and {Ridolfi}, A. and {Abbate}, F. and {Bailes}, M. and {Possenti}, A. and {Geyer}, M. and {Manchester}, R.~N. and {Kramer}, M. and {Freire}, P.~C.~C. and {Burgay}, M. and {Buchner}, S. and {Camilo}, F.},
        title = "{Timing of Millisecond Pulsars in NGC 6752. III. On the Presence of Nonluminous Matter in the Cluster's Core}",
      journal = {\apj},
         year = 2024,
        month = sep,
       volume = {972},
       number = {2},
          eid = {198},
        pages = {198},
          doi = {10.3847/1538-4357/ad5e74},
archivePrefix = {arXiv},
       eprint = {2407.03271},
 primaryClass = {astro-ph.HE},
       adsurl = {https://ui.adsabs.harvard.edu/abs/2024ApJ...972..198C}
}

@ARTICLE{2017MNRAS.468.2114P,
       author = {{Perera}, B.~B.~P. and {Stappers}, B.~W. and {Lyne}, A.~G. and {Bassa}, C.~G. and {Cognard}, I. and {Guillemot}, L. and {Kramer}, M. and {Theureau}, G. and {Desvignes}, G.},
        title = "{Evidence for an intermediate-mass black hole in the globular cluster NGC 6624}",
      journal = {\mnras},
         year = 2017,
        month = jun,
       volume = {468},
       number = {2},
        pages = {2114-2127},
          doi = {10.1093/mnras/stx501},
archivePrefix = {arXiv},
       eprint = {1705.01612},
 primaryClass = {astro-ph.HE},
       adsurl = {https://ui.adsabs.harvard.edu/abs/2017MNRAS.468.2114P}
}

@ARTICLE{2017MNRAS.471..857F,
       author = {{Freire}, P.~C.~C. and {Ridolfi}, A. and {Kramer}, M. and {Jordan}, C. and {Manchester}, R.~N. and {Torne}, P. and {Sarkissian}, J. and {Heinke}, C.~O. and {D'Amico}, N. and {Camilo}, F. and {Lorimer}, D.~R. and {Lyne}, A.~G.},
        title = "{Long-term observations of the pulsars in 47 Tucanae - II. Proper motions, accelerations and jerks}",
      journal = {\mnras},
         year = 2017,
        month = oct,
       volume = {471},
       number = {1},
        pages = {857-876},
          doi = {10.1093/mnras/stx1533},
archivePrefix = {arXiv},
       eprint = {1706.04908},
 primaryClass = {astro-ph.HE},
       adsurl = {https://ui.adsabs.harvard.edu/abs/2017MNRAS.471..857F}
}

@INPROCEEDINGS{1993ASPC...50..141P,
       author = {{Phinney}, E.~S.},
        title = "{Pulsars as Probes of Globular Cluster Dynamics}",
    booktitle = {Structure and Dynamics of Globular Clusters},
         year = 1993,
       editor = {{Djorgovski}, S.~G. and {Meylan}, Georges},
       series = {Astronomical Society of the Pacific Conference Series},
       volume = {50},
        month = jan,
        pages = {141},
       adsurl = {https://ui.adsabs.harvard.edu/abs/1993ASPC...50..141P}
}

@ARTICLE{2019SCPMA..6259502J,
       author = {{Jiang}, Peng and {Yue}, YouLing and {Gan}, HengQian and {Yao}, Rui and {Li}, Hui and {Pan}, GaoFeng and {Sun}, JingHai and {Yu}, DongJun and {Liu}, HongFei and {Tang}, NingYu and {Qian}, Lei and {Lu}, JiGuang and {Yan}, Jun and {Peng}, Bo and {Zhang}, ShuXin and {Wang}, QiMing and {Li}, Qi and {Li}, Di and {FAST Collaboration}},
        title = "{Commissioning progress of the FAST}",
      journal = {Science China Physics, Mechanics, and Astronomy},
         year = 2019,
        month = may,
       volume = {62},
       number = {5},
          eid = {959502},
        pages = {959502},
          doi = {10.1007/s11433-018-9376-1},
archivePrefix = {arXiv},
       eprint = {1903.06324},
 primaryClass = {astro-ph.IM},
       adsurl = {https://ui.adsabs.harvard.edu/abs/2019SCPMA..6259502J}
}

@ARTICLE{2012AR&T....9..237V,
       author = {{van Straten}, Willem and {Demorest}, Paul and {Oslowski}, Stefan},
        title = "{Pulsar Data Analysis with PSRCHIVE}",
      journal = {Astronomical Research and Technology},
         year = 2012,
        month = jul,
       volume = {9},
       number = {3},
        pages = {237-256},
          doi = {10.48550/arXiv.1205.6276},
archivePrefix = {arXiv},
       eprint = {1205.6276},
 primaryClass = {astro-ph.IM},
       adsurl = {https://ui.adsabs.harvard.edu/abs/2012AR&T....9..237V}
}

@ARTICLE{2011PASA...28....1V,
       author = {{van Straten}, W. and {Bailes}, M.},
        title = "{DSPSR: Digital Signal Processing Software for Pulsar Astronomy}",
      journal = {\pasa},
         year = 2011,
        month = jan,
       volume = {28},
       number = {1},
        pages = {1-14},
          doi = {10.1071/AS10021},
archivePrefix = {arXiv},
       eprint = {1008.3973},
 primaryClass = {astro-ph.IM},
       adsurl = {https://ui.adsabs.harvard.edu/abs/2011PASA...28....1V}
}

@ARTICLE{2018ApJ...863L..13A,
       author = {{Andersen}, Bridget C. and {Ransom}, Scott M.},
        title = "{A Fourier Domain {\textquotedblleft}Jerk{\textquotedblright} Search for Binary Pulsars}",
      journal = {\apjl},
         year = 2018,
        month = aug,
       volume = {863},
       number = {1},
          eid = {L13},
        pages = {L13},
          doi = {10.3847/2041-8213/aad59f},
archivePrefix = {arXiv},
       eprint = {1807.07900},
 primaryClass = {astro-ph.HE},
       adsurl = {https://ui.adsabs.harvard.edu/abs/2018ApJ...863L..13A}
}

@ARTICLE{2024Sci...383..275B,
       author = {{Barr}, Ewan D. and {Dutta}, Arunima and {Freire}, Paulo C.~C. and {Cadelano}, Mario and {Gautam}, Tasha and {Kramer}, Michael and {Pallanca}, Cristina and {Ransom}, Scott M. and {Ridolfi}, Alessandro and {Stappers}, Benjamin W. and {Tauris}, Thomas M. and {Venkatraman Krishnan}, Vivek and {Wex}, Norbert and {Bailes}, Matthew and {Behrend}, Jan and {Buchner}, Sarah and {Burgay}, Marta and {Chen}, Weiwei and {Champion}, David J. and {Chen}, C. -H. Rosie and {Corongiu}, Alessandro and {Geyer}, Marisa and {Men}, Y.~P. and {Padmanabh}, Prajwal Voraganti and {Possenti}, Andrea},
        title = "{A pulsar in a binary with a compact object in the mass gap between neutron stars and black holes}",
      journal = {Science},
         year = 2024,
        month = jan,
       volume = {383},
       number = {6680},
        pages = {275-279},
          doi = {10.1126/science.adg3005},
archivePrefix = {arXiv},
       eprint = {2401.09872},
 primaryClass = {astro-ph.HE},
       adsurl = {https://ui.adsabs.harvard.edu/abs/2024Sci...383..275B}
}

@ARTICLE{1985ApJ...294L..25D,
       author = {{Dewey}, R.~J. and {Taylor}, J.~H. and {Weisberg}, J.~M. and {Stokes}, G.~H.},
        title = "{A search for low-luminosity pulsars.}",
      journal = {\apjl},
         year = 1985,
        month = jul,
       volume = {294},
        pages = {L25-L29},
          doi = {10.1086/184502},
       adsurl = {https://ui.adsabs.harvard.edu/abs/1985ApJ...294L..25D}
}

@ARTICLE{2022MNRAS.511.5964Z,
       author = {{Zhao}, Jiaqi and {Heinke}, Craig O.},
        title = "{A census of X-ray millisecond pulsars in globular clusters}",
      journal = {\mnras},
         year = 2022,
        month = apr,
       volume = {511},
       number = {4},
        pages = {5964-5983},
          doi = {10.1093/mnras/stac442},
archivePrefix = {arXiv},
       eprint = {2202.07040},
 primaryClass = {astro-ph.HE},
       adsurl = {https://ui.adsabs.harvard.edu/abs/2022MNRAS.511.5964Z}
}

@ARTICLE{2002AJ....124.1788R,
       author = {{Ransom}, Scott M. and {Eikenberry}, Stephen S. and {Middleditch}, John},
        title = "{Fourier Techniques for Very Long Astrophysical Time-Series Analysis}",
      journal = {\aj},
         year = 2002,
        month = sep,
       volume = {124},
       number = {3},
        pages = {1788-1809},
          doi = {10.1086/342285},
archivePrefix = {arXiv},
       eprint = {astro-ph/0204349},
 primaryClass = {astro-ph},
       adsurl = {https://ui.adsabs.harvard.edu/abs/2002AJ....124.1788R}
}

@ARTICLE{2004PASA...21..302H,
       author = {{Hotan}, A.~W. and {van Straten}, W. and {Manchester}, R.~N.},
        title = "{PSRCHIVE and PSRFITS: An Open Approach to Radio Pulsar Data Storage and Analysis}",
      journal = {\pasa},
         year = 2004,
        month = jan,
       volume = {21},
       number = {3},
        pages = {302-309},
          doi = {10.1071/AS04022},
archivePrefix = {arXiv},
       eprint = {astro-ph/0404549},
 primaryClass = {astro-ph},
       adsurl = {https://ui.adsabs.harvard.edu/abs/2004PASA...21..302H}
}

@MISC{2015ascl.soft09002N,
       author = {{Nice}, D. and {Demorest}, P. and {Stairs}, I. and {Manchester}, R. and {Taylor}, J. and {Peters}, W. and {Weisberg}, J. and {Irwin}, A. and {Wex}, N. and {Huang}, Y.},
        title = "{Tempo: Pulsar timing data analysis}",
 howpublished = {Astrophysics Source Code Library, record ascl:1509.002},
         year = 2015,
        month = sep,
          eid = {ascl:1509.002},
        pages = {ascl:1509.002},
archivePrefix = {ascl},
       eprint = {1509.002},
       adsurl = {https://ui.adsabs.harvard.edu/abs/2015ascl.soft09002N}
}

@PHDTHESIS{2001PhDT.......123R,
       author = {{Ransom}, Scott Mitchell},
        title = "{New search techniques for binary pulsars}",
       school = {Harvard University, Massachusetts},
         year = 2001,
        month = jan,
       adsurl = {https://ui.adsabs.harvard.edu/abs/2001PhDT.......123R}
}

@ARTICLE{2007ApJ...670..363H,
       author = {{Hessels}, J.~W.~T. and {Ransom}, S.~M. and {Stairs}, I.~H. and {Kaspi}, V.~M. and {Freire}, P.~C.~C.},
        title = "{A 1.4 GHz Arecibo Survey for Pulsars in Globular Clusters}",
      journal = {\apj},
         year = 2007,
        month = nov,
       volume = {670},
       number = {1},
        pages = {363-378},
          doi = {10.1086/521780},
archivePrefix = {arXiv},
       eprint = {0707.1602},
 primaryClass = {astro-ph},
       adsurl = {https://ui.adsabs.harvard.edu/abs/2007ApJ...670..363H}
}

@ARTICLE{2011MNRAS.418..477B,
       author = {{Bagchi}, Manjari and {Lorimer}, D.~R. and {Chennamangalam}, Jayanth},
        title = "{Luminosities of recycled radio pulsars in globular clusters}",
      journal = {\mnras},
         year = 2011,
        month = nov,
       volume = {418},
       number = {1},
        pages = {477-489},
          doi = {10.1111/j.1365-2966.2011.19498.x},
archivePrefix = {arXiv},
       eprint = {1107.4521},
 primaryClass = {astro-ph.SR},
       adsurl = {https://ui.adsabs.harvard.edu/abs/2011MNRAS.418..477B}
}

@ARTICLE{1996MNRAS.282.1064H,
       author = {{Heggie}, Douglas C. and {Rasio}, Frederic A.},
        title = "{The Effect of Encounters on the Eccentricity of Binaries in Clusters}",
      journal = {\mnras},
         year = 1996,
        month = oct,
       volume = {282},
       number = {3},
        pages = {1064-1084},
          doi = {10.1093/mnras/282.3.1064},
archivePrefix = {arXiv},
       eprint = {astro-ph/9506082},
 primaryClass = {astro-ph},
       adsurl = {https://ui.adsabs.harvard.edu/abs/1996MNRAS.282.1064H}
}

@ARTICLE{2005Sci...307..892R,
       author = {{Ransom}, Scott M. and {Hessels}, Jason W.~T. and {Stairs}, Ingrid H. and {Freire}, Paulo C.~C. and {Camilo}, Fernando and {Kaspi}, Victoria M. and {Kaplan}, David L.},
        title = "{Twenty-One Millisecond Pulsars in Terzan 5 Using the Green Bank Telescope}",
      journal = {Science},
         year = 2005,
        month = feb,
       volume = {307},
       number = {5711},
        pages = {892-896},
          doi = {10.1126/science.1108632},
archivePrefix = {arXiv},
       eprint = {astro-ph/0501230},
 primaryClass = {astro-ph},
       adsurl = {https://ui.adsabs.harvard.edu/abs/2005Sci...307..892R}
}

@ARTICLE{1992RSPTA.341..117T,
       author = {{Taylor}, J.~H.},
        title = "{Pulsar Timing and Relativistic Gravity}",
      journal = {Philosophical Transactions of the Royal Society of London Series A},
         year = 1992,
        month = oct,
       volume = {341},
       number = {1660},
        pages = {117-134},
          doi = {10.1098/rsta.1992.0088},
       adsurl = {https://ui.adsabs.harvard.edu/abs/1992RSPTA.341..117T}
}

@ARTICLE{1992RSPTA.341...39P,
       author = {{Phinney}, E.~S.},
        title = "{Pulsars as Probes of Newtonian Dynamical Systems}",
      journal = {Philosophical Transactions of the Royal Society of London Series A},
         year = 1992,
        month = oct,
       volume = {341},
       number = {1660},
        pages = {39-75},
          doi = {10.1098/rsta.1992.0084},
       adsurl = {https://ui.adsabs.harvard.edu/abs/1992RSPTA.341...39P}
}

@ARTICLE{1962AJ.....67..471K,
       author = {{King}, Ivan},
        title = "{The structure of star clusters. I. an empirical density law}",
      journal = {\aj},
         year = 1962,
        month = oct,
       volume = {67},
        pages = {471},
          doi = {10.1086/108756},
       adsurl = {https://ui.adsabs.harvard.edu/abs/1962AJ.....67..471K}
}

@ARTICLE{2005ApJ...621..959F,
       author = {{Freire}, Paulo C.~C. and {Hessels}, Jason W.~T. and {Nice}, David J. and {Ransom}, Scott M. and {Lorimer}, Duncan R. and {Stairs}, Ingrid H.},
        title = "{The Millisecond Pulsars in NGC 6760}",
      journal = {\apj},
         year = 2005,
        month = mar,
       volume = {621},
       number = {2},
        pages = {959-965},
          doi = {10.1086/427748},
archivePrefix = {arXiv},
       eprint = {astro-ph/0411160},
 primaryClass = {astro-ph},
       adsurl = {https://ui.adsabs.harvard.edu/abs/2005ApJ...621..959F}
}

@ARTICLE{2020ApJ...892L...6P,
       author = {{Pan}, Zhichen and {Ransom}, Scott M. and {Lorimer}, Duncan R. and {Fiore}, William C. and {Qian}, Lei and {Wang}, Lin and {Stappers}, Benjamin W. and {Hobbs}, George and {Zhu}, Weiwei and {Yue}, Youling and {Wang}, Pei and {Lu}, Jiguang and {Liu}, Kuo and {Peng}, Bo and {Zhang}, Lei and {Li}, Di},
        title = "{The FAST Discovery of an Eclipsing Binary Millisecond Pulsar in the Globular Cluster M92 (NGC 6341)}",
      journal = {\apjl},
         year = 2020,
        month = mar,
       volume = {892},
       number = {1},
          eid = {L6},
        pages = {L6},
          doi = {10.3847/2041-8213/ab799d},
archivePrefix = {arXiv},
       eprint = {2002.10337},
 primaryClass = {astro-ph.HE},
       adsurl = {https://ui.adsabs.harvard.edu/abs/2020ApJ...892L...6P}
}

@ARTICLE{2001MNRAS.326..274L,
       author = {{Lange}, Ch. and {Camilo}, F. and {Wex}, N. and {Kramer}, M. and {Backer}, D.~C. and {Lyne}, A.~G. and {Doroshenko}, O.},
        title = "{Precision timing measurements of PSR J1012+5307}",
      journal = {\mnras},
         year = 2001,
        month = sep,
       volume = {326},
       number = {1},
        pages = {274-282},
          doi = {10.1046/j.1365-8711.2001.04606.x},
archivePrefix = {arXiv},
       eprint = {astro-ph/0102309},
 primaryClass = {astro-ph},
       adsurl = {https://ui.adsabs.harvard.edu/abs/2001MNRAS.326..274L}
}

@ARTICLE{1993ApJ...415..631S,
       author = {{Sigurdsson}, Steinn and {Phinney}, E.~S.},
        title = "{Binary--Single Star Interactions in Globular Clusters}",
      journal = {\apj},
         year = 1993,
        month = oct,
       volume = {415},
        pages = {631},
          doi = {10.1086/173190},
       adsurl = {https://ui.adsabs.harvard.edu/abs/1993ApJ...415..631S}
}

@ARTICLE{1999A&A...350..928T,
       author = {{Tauris}, Thomas M. and {Savonije}, Gerrit J.},
        title = "{Formation of millisecond pulsars. I. Evolution of low-mass X-ray binaries with P\_orb> 2 days}",
      journal = {\aap},
         year = 1999,
        month = oct,
       volume = {350},
        pages = {928-944},
          doi = {10.48550/arXiv.astro-ph/9909147},
archivePrefix = {arXiv},
       eprint = {astro-ph/9909147},
 primaryClass = {astro-ph},
       adsurl = {https://ui.adsabs.harvard.edu/abs/1999A&A...350..928T}
}

@ARTICLE{2017ApJ...835...29Y,
       author = {{Yao}, J.~M. and {Manchester}, R.~N. and {Wang}, N.},
        title = "{A New Electron-density Model for Estimation of Pulsar and FRB Distances}",
      journal = {\apj},
         year = 2017,
        month = jan,
       volume = {835},
       number = {1},
          eid = {29},
        pages = {29},
          doi = {10.3847/1538-4357/835/1/29},
archivePrefix = {arXiv},
       eprint = {1610.09448},
 primaryClass = {astro-ph.GA},
       adsurl = {https://ui.adsabs.harvard.edu/abs/2017ApJ...835...29Y}
}

@ARTICLE{2021MNRAS.504.1407R,
       author = {{Ridolfi}, A. and {Gautam}, T. and {Freire}, P.~C.~C. and {Ransom}, S.~M. and {Buchner}, S.~J. and {Possenti}, A. and {Venkatraman Krishnan}, V. and {Bailes}, M. and {Kramer}, M. and {Stappers}, B.~W. and {Abbate}, F. and {Barr}, E.~D. and {Burgay}, M. and {Camilo}, F. and {Corongiu}, A. and {Jameson}, A. and {Padmanabh}, P.~V. and {Vleeschower}, L. and {Champion}, D.~J. and {Chen}, W. and {Geyer}, M. and {Karastergiou}, A. and {Karuppusamy}, R. and {Parthasarathy}, A. and {Reardon}, D.~J. and {Serylak}, M. and {Shannon}, R.~M. and {Spiewak}, R.},
        title = "{Eight new millisecond pulsars from the first MeerKAT globular cluster census}",
      journal = {\mnras},
         year = 2021,
        month = jun,
       volume = {504},
       number = {1},
        pages = {1407-1426},
          doi = {10.1093/mnras/stab790},
archivePrefix = {arXiv},
       eprint = {2103.04800},
 primaryClass = {astro-ph.HE},
       adsurl = {https://ui.adsabs.harvard.edu/abs/2021MNRAS.504.1407R}
}

@INPROCEEDINGS{2008IAUS..246..291R,
       author = {{Ransom}, Scott M.},
        title = "{Pulsars in Globular Clusters}",
    booktitle = {Dynamical Evolution of Dense Stellar Systems},
         year = 2008,
       editor = {{Vesperini}, Enrico and {Giersz}, Mirek and {Sills}, Alison},
       series = {IAU Symposium},
       volume = {246},
        month = may,
        pages = {291-300},
          doi = {10.1017/S1743921308015810},
       adsurl = {https://ui.adsabs.harvard.edu/abs/2008IAUS..246..291R}
}

@ARTICLE{2007AJ....133.2787P,
       author = {{Paust}, Nathaniel E.~Q. and {Chaboyer}, Brian and {Sarajedini}, Ata},
        title = "{BVI Photometry and the Luminosity Functions of the Globular Cluster M92}",
      journal = {\aj},
         year = 2007,
        month = jun,
       volume = {133},
       number = {6},
        pages = {2787-2798},
          doi = {10.1086/513511},
archivePrefix = {arXiv},
       eprint = {astro-ph/0703167},
 primaryClass = {astro-ph},
       adsurl = {https://ui.adsabs.harvard.edu/abs/2007AJ....133.2787P}
}

@ARTICLE{2025arXiv251211058D,
       author = {{Das}, Jyotirmoy and {Roy}, Jayanta and {Freire}, Paulo C.~C. and {Ransom}, Scott M and {Bhattacharyya}, Bhaswati and {Ad{\'a}mek}, Karel and {Armour}, Wes and {Kudale}, Sanjay and {Muley}, Mekhala V.},
        title = "{Globular Clusters GMRT Pulsar Search (GCGPS) II: Discovery of five MSPs in M69 and M70}",
      journal = {arXiv e-prints},
         year = 2025,
        month = dec,
          eid = {arXiv:2512.11058},
        pages = {arXiv:2512.11058},
          doi = {10.48550/arXiv.2512.11058},
archivePrefix = {arXiv},
       eprint = {2512.11058},
 primaryClass = {astro-ph.HE},
       adsurl = {https://ui.adsabs.harvard.edu/abs/2025arXiv251211058D}
}

@ARTICLE{2025ApJ...988..161D,
       author = {{Das}, Jyotirmoy and {Roy}, Jayanta and {Freire}, Paulo C.~C. and {Ransom}, Scott M. and {Bhattacharyya}, Bhaswati and {Ad{\'a}mek}, Karel and {Armour}, Wes and {Kudale}, Sanjay and {Muley}, Mekhala V.},
        title = "{Globular Clusters GMRT Pulsar Search (GCGPS). I. Survey Description, Discovery and Timing of the First Pulsar in NGC6093 (M80)}",
      journal = {\apj},
         year = 2025,
        month = aug,
       volume = {988},
       number = {2},
          eid = {161},
        pages = {161},
          doi = {10.3847/1538-4357/ade052},
archivePrefix = {arXiv},
       eprint = {2502.09154},
 primaryClass = {astro-ph.HE},
       adsurl = {https://ui.adsabs.harvard.edu/abs/2025ApJ...988..161D}
}

@ARTICLE{2023RAA....23e5012Y,
       author = {{Yin}, De-Jiang and {Zhang}, Li-Yun and {Li}, Bao-Da and {Li}, Ming-Hui and {Qian}, Lei and {Pan}, Zhichen},
        title = "{The Analyses of Globular Cluster Pulsars and Their Detection Efficiency}",
      journal = {Research in Astronomy and Astrophysics},
         year = 2023,
        month = may,
       volume = {23},
       number = {5},
          eid = {055012},
        pages = {055012},
          doi = {10.1088/1674-4527/acc37e},
archivePrefix = {arXiv},
       eprint = {2303.09710},
 primaryClass = {astro-ph.HE},
       adsurl = {https://ui.adsabs.harvard.edu/abs/2023RAA....23e5012Y}
}

@ARTICLE{2013MNRAS.436.3720T,
       author = {{Turk}, P.~J. and {Lorimer}, D.~R.},
        title = "{An empirical Bayesian analysis applied to the globular cluster pulsar population}",
      journal = {\mnras},
         year = 2013,
        month = dec,
       volume = {436},
       number = {4},
        pages = {3720-3726},
          doi = {10.1093/mnras/stt1850},
archivePrefix = {arXiv},
       eprint = {1309.7317},
 primaryClass = {astro-ph.GA},
       adsurl = {https://ui.adsabs.harvard.edu/abs/2013MNRAS.436.3720T}
}

@ARTICLE{2013MNRAS.431..874C,
       author = {{Chennamangalam}, Jayanth and {Lorimer}, D.~R. and {Mandel}, Ilya and {Bagchi}, Manjari},
        title = "{Constraining the luminosity function parameters and population size of radio pulsars in globular clusters}",
      journal = {\mnras},
         year = 2013,
        month = may,
       volume = {431},
       number = {1},
        pages = {874-881},
          doi = {10.1093/mnras/stt205},
archivePrefix = {arXiv},
       eprint = {1207.5732},
 primaryClass = {astro-ph.SR},
       adsurl = {https://ui.adsabs.harvard.edu/abs/2013MNRAS.431..874C}
}

@ARTICLE{2010arXiv1012.3224H,
       author = {{Harris}, William E.},
        title = "{A New Catalog of Globular Clusters in the Milky Way}",
      journal = {arXiv e-prints},
         year = 2010,
        month = dec,
          eid = {arXiv:1012.3224},
        pages = {arXiv:1012.3224},
          doi = {10.48550/arXiv.1012.3224},
archivePrefix = {arXiv},
       eprint = {1012.3224},
 primaryClass = {astro-ph.GA},
       adsurl = {https://ui.adsabs.harvard.edu/abs/2010arXiv1012.3224H}
}

@ARTICLE{1995ApJ...445L.133R,
       author = {{Rasio}, Frederic A. and {Heggie}, Douglas C.},
        title = "{The Orbital Eccentricities of Binary Millisecond Pulsars in Globular Clusters}",
      journal = {\apjl},
         year = 1995,
        month = jun,
       volume = {445},
        pages = {L133},
          doi = {10.1086/187907},
archivePrefix = {arXiv},
       eprint = {astro-ph/9502105},
 primaryClass = {astro-ph},
       adsurl = {https://ui.adsabs.harvard.edu/abs/1995ApJ...445L.133R}
}

@ARTICLE{2023MNRAS.525.4167O,
       author = {{Oh}, Kwangmin and {Hui}, C.~Y. and {Hong}, Jongsuk and {Takata}, J. and {Kong}, A.~K.~H. and {Tam}, Pak-Hin Thomas and {Li}, Kwan-Lok and {Cheng}, K.~S.},
        title = "{Influences of dynamical disruptions on the evolution of pulsars in globular clusters}",
      journal = {\mnras},
         year = 2023,
        month = nov,
       volume = {525},
       number = {3},
        pages = {4167-4175},
          doi = {10.1093/mnras/stad2383},
archivePrefix = {arXiv},
       eprint = {2308.04920},
 primaryClass = {astro-ph.HE},
       adsurl = {https://ui.adsabs.harvard.edu/abs/2023MNRAS.525.4167O}
}

@ARTICLE{2014A&A...561A..11V,
       author = {{Verbunt}, Frank and {Freire}, Paulo C.~C.},
        title = "{On the disruption of pulsar and X-ray binar ies in globular clusters}",
      journal = {\aap},
         year = 2014,
        month = jan,
       volume = {561},
          eid = {A11},
        pages = {A11},
          doi = {10.1051/0004-6361/201321177},
archivePrefix = {arXiv},
       eprint = {1310.4669},
 primaryClass = {astro-ph.SR},
       adsurl = {https://ui.adsabs.harvard.edu/abs/2014A&A...561A..11V}
}

@ARTICLE{2021ApJ...915L..28P,
       author = {{Pan}, Zhichen and {Qian}, Lei and {Ma}, Xiaoyun and {Liu}, Kuo and {Wang}, Lin and {Luo}, Jintao and {Yan}, Zhen and {Ransom}, Scott and {Lorimer}, Duncan and {Li}, Di and {Jiang}, Peng},
        title = "{FAST Globular Cluster Pulsar Survey: Twenty-four Pulsars Discovered in 15 Globular Clusters}",
      journal = {\apjl},
         year = 2021,
        month = jul,
       volume = {915},
       number = {2},
          eid = {L28},
        pages = {L28},
          doi = {10.3847/2041-8213/ac0bbd},
archivePrefix = {arXiv},
       eprint = {2106.08559},
 primaryClass = {astro-ph.HE},
       adsurl = {https://ui.adsabs.harvard.edu/abs/2021ApJ...915L..28P}
}

@ARTICLE{2011ApJ...734...89L,
       author = {{Lynch}, Ryan S. and {Ransom}, Scott M. and {Freire}, Paulo C.~C. and {Stairs}, Ingrid H.},
        title = "{Six New Recycled Globular Cluster Pulsars Discovered with the Green Bank Telescope}",
      journal = {\apj},
         year = 2011,
        month = jun,
       volume = {734},
       number = {2},
          eid = {89},
        pages = {89},
          doi = {10.1088/0004-637X/734/2/89},
archivePrefix = {arXiv},
       eprint = {1101.1467},
 primaryClass = {astro-ph.SR},
       adsurl = {https://ui.adsabs.harvard.edu/abs/2011ApJ...734...89L}
}
\bibliographystyle{aasjournalv7}

% \appendix

\end{document}